\documentclass[journal=jacsat,manuscript=article]{achemso}
\usepackage[version=3]{mhchem} % Formula subscripts using \ce{}

\usepackage{amsmath}
\usepackage{amssymb}
\usepackage{graphicx}% Include figure files
\usepackage{dcolumn}% Align table columns on decimal point
\usepackage{bm}% bold math
\usepackage{mathptmx}
\usepackage{etoolbox}
\usepackage{xcolor}
\usepackage{subcaption}
\usepackage{physics}
\usepackage[normalem]{ulem}

\title{Simple fluctuations in simple glass formers}

\author{Corentin C. L. Laudicina}
\affiliation{Soft Matter and Biological Physics, Department of Applied Physics, Eindhoven University of Technology, P.O.\,Box 513, 5600 MB Eindhoven, The Netherlands}
\email{c.c.l.laudicina@tue.nl}
\author{Patrick Charbonneau}
\author{Yi Hu}
\affiliation{Department of Chemistry, Duke University, Durham, North Carolina 27708, USA}
\author{Liesbeth M. C. Janssen}
\affiliation{Soft Matter and Biological Physics, Department of Applied Physics, Eindhoven University of Technology, P.O.\,Box 513, 5600 MB Eindhoven, The Netherlands}
\author{Peter K. Morse}
\affiliation{Department of Chemistry, Department of Physics, and Princeton Institute of Materials, Princeton University, Princeton, New Jersey 08544, USA}
\author{Ilian Pihlajamaa}
\affiliation{Soft Matter and Biological Physics, Department of Applied Physics, Eindhoven University of Technology, P.O.\,Box 513, 5600 MB Eindhoven, The Netherlands}

\author{Grzegorz Szamel}
\affiliation{Department of Chemistry, Colorado State University, Fort Collins, Colorado 80523, USA}
\begin{document}

\begin{abstract}
Critical single-particle fluctuations associated with particle displacements are inherent to simple glass-forming liquids in the limit of large dimensions and leave a pseudo-critical trace across all finite dimensions. This characteristic could serve as a crucial test for distinguishing between theories of glass formation. We here examine these critical fluctuations, as captured by the well-established non-Gaussian parameter, within both mode-coupling theory (MCT) and dynamical mean-field theory (DMFT) across dimensions for hard sphere liquids and for the minimally structured Mari--Kurchan model. We establish general scaling laws relevant to any liquid dynamics theory in large dimensions and show that the dimensional scalings predicted by MCT are inconsistent with those from DMFT. Simulation results for hard spheres in moderately high dimensions align with the DMFT scenario, reinforcing the relevance of mean-field theory for capturing glass physics in finite dimensions. We identify potential adjustments to MCT to account for certain mean-field physics. Our findings also highlight that local structure and spatial dimensionality can affect single-particle critical fluctuations in non-trivial ways.
\end{abstract}

\maketitle

\section{\label{sec:level1}Introduction}

A key challenge with the microscopic understanding of glasses is that a plethora of distinct yet seemingly incompatible physical theories describe their formation from supercooled liquids.\cite{berthier2011} The universal phenomenon of glass formation is characterized by the sudden growth of the viscosity (or equivalently the structural relaxation time) of a liquid within a small temperature or density change past the melting point, eventually leading to the emergence of glass rigidity and the existence of an amorphous solid. All theories of the glass transition -- necessarily -- capture this phenomenology reasonably well.\cite{tarjus2011overview} While there is no clear consensus on the subject, there is nevertheless strong evidence that glassy behavior is governed by a genuine, although potentially unreachable thermodynamic phase transition,\cite{berthier2017,guiselin2022statistical} itself preceded by an avoided dynamical transition.\cite{kirkpatrick1987, biroli2012random,parisi2019, biroli2023rfot} As a result, the dynamics in the supercooled regime are expected to be governed by a complex and potentially non-perturbative pseudo-critical regime. From a statistical physics standpoint, a particularly stringent test of theoretical proposals comes from their description of critical fluctuations. Just as a theory of ferromagnetism must predict a diverging susceptibility at the Curie point, a robust theory of glass formation should accurately account for critical fluctuations (or remnants of thereof), which in the context of the glass formation are generally understood to correspond to dynamical heterogeneity. Interestingly, significantly different predictions of this phenomenon have been made.\cite{sillescu1999heterogeneity, berthier2011}

The growth of dynamical heterogeneity with the structural relaxation time $\tau_\alpha$ was first noted nearly three decades ago.\cite{kob1997, donati1998stringlike, donati1999spatial} The simplest aspects of such heterogeneity are captured by a single-particle observable, the non-Gaussian parameter (NGP), commonly denoted $\alpha_2(t)$.\cite{footnote1} Other quantifiers of dynamical heterogeneity such as three- and four-point functions are typically used to capture the spatial extent of such heterogeneity.\cite{glotzer2000time, donati2002theory,berthier2007spontaneousII, flenner2011analysis} The NGP, which by definition quantifies deviations from Gaussian-distributed particle displacements, vanishes for non-interacting particles with either Brownian or Newtonian dynamics as well as for interacting particles in the diffusive regime. It is therefore zero at both short and long times in all (monodisperse) systems interacting with pair-wise additive interactions, i.e., simple liquids.
Its non-trivial behavior at intermediate times can be variously interpreted. In particular, the existence of a non-zero NGP has been argued to evince the existence of particles existing in two subpopulations: transiently mobile and transiently immobile particles.\cite{berthier2011, adhikari2021} 
Interpretations of the sort, while intuitively useful, however do not necessarily lead to specific quantitative predictions that could be tested against $\alpha_2(t)$ measured in computer simulations or experiments. 

In this context, it is particularly interesting to consider the mode-coupling theory (MCT) of the glass transition,\cite{bengtzelius1984dynamics, leutheusser1984dynamical, gotze2009complex} which is one of the few approaches to make near-quantitative predictions of supercooled liquid dynamics. Although the MCT description has been extensively studied in the past with applications to a wide variety of realistic structural glass formers (see Ref.~\citenum{gotze1999recent} for a historical review and Ref.~\citenum{janssen2018mode} for a more recent one), no clear and systematic test of MCT with regards to simple dynamical fluctuations such as the NGP has yet been considered. Because MCT has also been considered in the limit of large spatial dimensions where mean-field (and hence fluctuation-free) behavior is expected to become dominant,\cite{ikeda2010,schmid2010} it also has the potential to provide insight about the onset of the mean-field--like regime.

The dynamical mean-field theory (DMFT) of glasses\cite{maimbourg2016solution} also offers an interesting perspective on such fluctuations. The DMFT description, which is %quantitatively 
exact in the limit of high spatial dimension $d\rightarrow\infty$, suggests that 
$\alpha_2(t)$ is a perturbative correction in $1/d$, given that purely Gaussian dynamics is expected in that limit.\cite{biroli2022} Physically, this framing associates the NGP with growing correlations among the directions along which a single particle moves as the liquid turns sluggish. Unfortunately, an explicit expression for $\alpha_2(t)$ in the fluid regime has not been derived yet. 
Recently, however, a way around this issue has been advanced. By simulating a minimally-structured liquid model in finite $d$ and
then rescaling the finite $d$ results, one can indeed validate the DMFT scenario and even infer some of its (putative) finite-$d$ corrections.\cite{charbonneau2024dynamics}

In this article, we consider $\alpha_2(t)$ calculated within MCT and DMFT, the only two approaches to offer (quasi-)quantitative predictions for the observable. Specifically, we investigate critical simple fluctuations in the hard sphere (HS) fluid, and the minimally structured Mari--Kurchan (MK) model\cite{mari2011dynamical}, which are known to belong to the same dynamical universality class in the large dimensional limit. We first analytically examine the dimensional scaling of the equations of motion from which the NGP can be derived. At low and intermediate densities, we then demonstrate that the results from numerical simulations generally align with the DMFT scenario, while MCT incorrectly captures the dimensional trend but does accurately predict some associated scaling relations. These findings generally validate the relevance of the dynamical mean-field theory of glasses for finite-$d$ systems. Surprisingly, NGP-captured critical fluctuations in low-$d$ glass formers are much suppressed relative to what one infers from the DMFT. This result allows us to claim that fluid structure might play a crucial role in suppressing the fluctuations that accompany dynamical arrest relative to the mean-field scenario. 

The rest of this article is structured as follows. First, we present general results regarding the NGP in the liquid state, focusing on the dimensional scaling of the memory kernels of the equations of motion for dynamical correlation functions, from which the NGP can be derived. Next, we explicitly compare the dimensional scalings of the critical behavior predicted by MCT for high dimensional hard spheres and the MK model. Finally, numerical MCT predictions for the NGP are obtained for both systems and then qualitatively and quantitatively compared with results from computer simulations.

\section{Dimensional Scalings of the Non-Gaussian Parameter} \label{sec:general_NGP}
We first derive general equations of motion for the moments of the displacements of a particle in a liquid of arbitrary dimension. This allows us to obtain explicit conditions for the correct dimensional scalings of the mean squared displacement, the non-Gaussian parameter, and related quantities. These conditions are later used as a benchmark for comparing simulation results and our mode-coupling study.

\subsection{General Equations of Motion}
\label{sec:generaleq}
Let us first consider the general properties of the NGP for a $d$-dimensional monatomic liquid composed of $N$ particles, defined as
    \begin{equation}
        \alpha_2(t ; d) = \frac{d}{d+2}\frac{\langle \Delta r^4(t ; d)\rangle}{\langle \Delta r^2(t ; d)\rangle^2} - 1
    \label{eq:NGP_def}
    \end{equation}
at time $t$, where $\langle \Delta r^2(t ; d)\rangle$, $\langle \Delta r^4(t ; d)\rangle$ are the mean squared displacement (MSD) and the mean quartic displacement (MQD), respectively. Angular brackets $\langle \cdot\rangle$ denote ensemble averaging. The MSD and the MQD correspond respectively to the second ($n=2$) and fourth ($n=4$) moments of the self part of the Van Hove function,
    \begin{equation}
        \langle \Delta r^n(t; d)\rangle \equiv \int_{\mathbb{R}^d}\mathrm{d}\boldsymbol{r} r^n G_s(r,t),
    \end{equation}
where $G_s(r,t) \equiv N^{-1}\langle \sum_{i=1}^N \delta(\boldsymbol{r}- \Delta \boldsymbol{r}_i(t))\rangle$ in which $\Delta \boldsymbol{r}_i(t)$ denotes the displacement vector of particle $i$ at time $t$.\cite{hansen2006} By translational and rotational invariance of the equilibrium liquid, $G_s(r,t)$ only depends on the modulus of the displacement vector $r \equiv |\boldsymbol{r}|$. The Fourier transform of the self Van Hove function further defines the self-intermediate scattering function 
    \begin{equation}
        \phi_s(q,t ; d) = \int_{\mathbb{R}^d}\mathrm{d}\boldsymbol{r} e^{-i\boldsymbol{q}\cdot\boldsymbol{r}}G_s(r,t),
    \label{eq:sISF_def}
    \end{equation}
where $q$ is magnitude of the wavevector $\boldsymbol{q}$ at which fluctuations are probed. This dynamical quantity can be measured from scattering experiments and is a common liquid-state observable.\cite{hansen2006} 
The self-intermediate scattering function can also be related to the MSD and the NGP:~\cite{rahman1964correlations, fuchs1998asymptotic} 
    \begin{equation}
    \begin{split}        
        \phi_s(q, t ; d) %&\sum_{n=0}^{\infty} \frac{(-1)^n}{(2n)!} \int_{\mathbb{R}^d} d\boldsymbol{r}\ (\boldsymbol{q}\cdot\boldsymbol{r})^{2n}G_s(r,t)  %\\
        =&\ e^{-\frac{q^2\langle \Delta r^2(t)\rangle}{2d}}\left[1 + \frac{\alpha_2(t ; d)}{2}\left(\frac{q^2\langle \Delta r^2(t ; d)\rangle}{2d} \right)^2  + \mathcal{O}(q^6)\right].
    \end{split}
    \label{eq:series_tagged_ISF}
    \end{equation}
For purely Gaussian dynamical processes, $\alpha_2(t ; d) = 0$, and all higher order terms in $q$ in the square brackets of Eq.~\eqref{eq:series_tagged_ISF} vanish. As a result, in the glass phase, where the MSD plateaus at some value $R$ (i.e. $\langle \Delta r^2(t\rightarrow\infty ; d)\rangle = R$), the non-ergodicity parameter (NEP) $\phi_s(q,t\rightarrow\infty ; d) \equiv \phi_s^{\infty}(q ; d)>0$ becomes a Gaussian function of $q$,
    \begin{equation}
        \phi_s^{\infty}(q ; d) = e^{-Rq^2 / 2d}.
    \label{eq:Gaussian_NEP}
    \end{equation}
In other words, ergodicity breaking of a Gaussian process in real space necessarily corresponds to a Gaussian non-ergodicity parameter, irrespective of the dimension of space.

We next consider how generally to obtain an equation of motion for $\alpha_2(t)$. Note that for the sake of simplicity, we only consider the behavior of a liquid in the overdamped dynamical regime, but the final expressions trivially generalize to liquids in the underdamped dynamical regime as well. For notational simplicity, we also drop the explicit $d$-dependence from the arguments of the MSD, MQD, and NGP, as well as those of related quantities. The starting point is a formally exact integro-differential equation for $\phi_s(q,t)$ derived using either projection operators\cite{mori1965transport, zwanzig2001nonequilibrium} or field-theoretic considerations,\cite{das1986fluctuating, miyazaki2005mode}
    \begin{equation}
        \gamma_s(q)\pdv{\phi_s(q,t)}{t} + \phi_s(q,t) + \int_0^t \mathrm{d}\tau m_s(q,t-\tau)\pdv{\phi_s(q,\tau)}{\tau} = 0,
    \label{eq:GLE_tagged_ISF}
    \end{equation}
where $\gamma_s(q)\equiv (q^2D_0)^{-1}$ sets the natural timescale for the decay of a single particle density fluctuation in free space with $D_0$ the bare diffusion constant, and the integral kernel $m_s(q,t)$ is commonly referred to as a memory term (or self-energy) due to its non-time local contribution. Inserting the series expansion of Eq.~\eqref{eq:sISF_def} in Eq.~\eqref{eq:GLE_tagged_ISF} and matching terms of equivalent order in $q$ then gives the equation of motion for the MSD
    \begin{equation}
        \dv{\langle\Delta r^2(t)\rangle}{t} + D_0\int_0^t\mathrm{d}\tau\ m_0(t-\tau)\dv{\langle \Delta r^2(\tau)\rangle}{\tau} = 2dD_0,
    \label{eq:general_eom_MSD}
    \end{equation}
and the MQD
    \begin{equation}
        \dv{\langle\Delta r^4(t)\rangle}{t} + D_0 \int_0^t \mathrm{d}\tau\ m_0(t-\tau)\dv{\langle \Delta r^4(\tau)\rangle}{\tau} = \mu(t ; d),
    \label{eq:general_eom_fourth}
    \end{equation}
where $m_0(t) \equiv \lim_{q\rightarrow0} q^2m_s(q,t)$. The inhomogeneous term in Eq.~\eqref{eq:general_eom_fourth} reads
    \begin{equation}
    \begin{split}
        \mu(t ; d) = &\ 4(d+2)D_0\langle \Delta r^2(t)\rangle + 2(d+2)\int_0^t\mathrm{d}\tau\  m_2(t-\tau)\dv{\langle \Delta r^2(\tau)\rangle}{\tau},
    \end{split}
    \end{equation}
where $m_2(t)\equiv \lim_{q\rightarrow 0}q^2\partial_q^2 m_s(q,t)$. The factors $q^2$ make $m_0(t)$ and $m_2(t)$ well-behaved. They are needed since the single-particle dynamics does not conserve the particle's momentum (due to molecular collisions with surrounding particles).\cite{gotze2009complex} If $m_0(t)$ and $m_2(t)$ are known, Eqs.~\eqref{eq:general_eom_MSD}--\eqref{eq:general_eom_fourth} can then be solved by direct integration, and the NGP extracted via Eq.~\eqref{eq:NGP_def}. We must note that in the case of DMFT (which has only been developed for the MSD), the memory kernel $m_0(t)$ instead consists of a self-consistently determined stochastic process,\cite{parisi2019} for which no explicit expression exists. This is in contrast with MCT which offers an explicit expression for the memory kernels in terms of structural observables,\cite{gotze2009complex} as further discussed below.

\subsection{Determination of Dimensional Scalings} %\label{sec:general_dim_scaling}
In order to assess the impact of dimension on the liquid dynamics, we next consider the $d$-dependent scaling of the formally exact equations Eqs.~\eqref{eq:GLE_tagged_ISF}--\eqref{eq:general_eom_fourth}. We here specifically posit the following scaling forms: $ \langle \Delta r^2(t)\rangle \sim d^{\alpha}$, $\langle \Delta r^4(t)\rangle \sim d^{\beta}$, and $\alpha_2(t) \sim d^{\varsigma}$ for the standard observables. One may also write $m_0(t) \sim d^{\upsilon}$, $m_2(t) \sim d^{\lambda}$ and $D_0 \sim d^{\eta}$. To facilitate the ensuing analysis, we consider the non-ergodic side of the state diagram to extract the long-time limits of the MSD and the MQD given by Eqs.~\eqref{eq:general_eom_MSD}--\eqref{eq:general_eom_fourth}, respectively. Throughout this work, we assume that ergodicity breaking does not modify dimensional scalings. We find that the long-time limit of the MSD reads
    \begin{equation}
        \langle \Delta r^2(t\rightarrow\infty) \rangle = \frac{2d}{m_0(t\rightarrow\infty)},
    \label{eq:long_time_MSD}
    \end{equation}
and similarly for the MQD
    \begin{equation}
        \langle \Delta r^4(t\rightarrow\infty) \rangle = \frac{2(d+2)\big[2 + m_2(t\rightarrow\infty)\big] \langle \Delta r^2(t\rightarrow\infty)\rangle}{m_0(t\rightarrow\infty)}.
    \label{eq:long_time_MQD}
    \end{equation}
Note that both expressions are expected to be independent of microscopic dynamics (here seen via their independence from the bare diffusion constant $D_0$). Simple power-counting arguments allow us to derive algebraic relations between the various exponents governing the asymptotic dimensional scaling. From Eq.~\eqref{eq:long_time_MSD}, one determines that $\alpha = 1-\upsilon$. Similarly, for the MQD we have
    \begin{equation}
        \beta = \begin{cases}
			1+\alpha-\upsilon & \text{if } \lambda < 0 \\
            1+\alpha-\upsilon+\lambda, & \text{if } \lambda > 0,
		 \end{cases}
    \end{equation}
and the NGP $\varsigma = \beta - 2\alpha$. Note that these power-counting arguments provide upper bounds for the dimensional scalings, and do not inform us about the magnitude of the numerical prefactors. 

We next recapitulate known results for the dynamics of liquids in the limit of large dimensions, in which the DMFT becomes exact. In this limit, it has been established that to study the kinetic arrest at the dynamical transition, an investigation of the dynamics for length scales of order $\sim \mathcal{O}(d^{-1/2})$ suffices.~\cite{maimbourg2016solution, parisi2019} This result naturally motivates us to express the MSD as $ \langle \Delta r^2(t)\rangle \equiv d^{\alpha}\times \hat{\Delta}_2(t)$ and similarly the MQD as $\langle \Delta r^4(t) \rangle \equiv d^{\beta} \times\hat{\Delta}_4(t)$ with $\alpha=-1$, $\beta=-2$ and where $\hat{\Delta}_{2,4}(t)$ should be finite as $d\rightarrow\infty$.\cite{maimbourg2016solution, liu2021dynamics} This constraint suffices to enforce (near) Gaussian statistics at large $d$ since we naively find that $\varsigma = 0$ from power counting arguments, and therefore that $\alpha_2(t ; d\rightarrow\infty) \sim d^0$. However, a more refined analysis reveals that the pre-factor vanishes in this case, and that the leading order contribution in fact gives $\varsigma = -1$.\cite{biroli2022} This scaling motivates interpreting the non-Gaussian dynamics as a perturbative $1/d$ correction, and hence that  $\alpha_2(t ; d) = d^{-1} \times\hat{\alpha}_2(t)$, with $\hat{\alpha}_2(t)$ a well-behaved function in the limit $d\rightarrow\infty$.\cite{biroli2022}

We can then use these scalings to infer those of the memory kernels introduced in the previous section. In particular, we find that $m_0(t) \equiv d^2 \times \hat{M}_0(t)$ and $m_2(t) \equiv d^0 \times \hat{M}_2(t)$ where $\hat{M}_0(t),\ \hat{M}_2(t)$ are finite as $d\rightarrow\infty$. The above scalings and the analysis of Eq.~\eqref{eq:general_eom_MSD} imply that non-trivial, glassy physics will only be observed if the bare diffusion constant is of order $D_0 \equiv d^{-2} \times \hat{D}_0$. This allows us to consider an additional important quantity, namely the long-time diffusion constant,
\begin{equation}
        D = D_0\left( 1 + D_0\int_0^{\infty}\mathrm{d}t\ m_0(t)\right)^{-1},
    \end{equation}
from which we conclude that $D = d^{\eta}\times \hat{D}$ with $\eta = -2$ and $\hat{D}$ finite at large $d$, in agreement with known results.\cite{maimbourg2016solution,manacorda2020} 
%Note that because a theory for liquid dynamics necessarily obeys equations of the form of Eqs.~\eqref{eq:GLE_tagged_ISF}, \eqref{eq:general_eom_MSD} and \eqref{eq:general_eom_fourth},
The above discussion imposes general constraints on the form of the memory kernels for liquids in the limit $d\rightarrow\infty$. 

\section{Dimensional Scaling of MCT for the Hard-Sphere Universality Class}
\label{sec:MCT_scaling}

With the natural dimensional scalings %from Sec.~\ref{sec:general_NGP}
in hand, this section examines the $d$-dependent scaling of the microscopic MCT equations. Although originally designed to investigate the collective relaxation of density modes in supercooled liquids through predictions of the intermediate scattering function $\phi(q,t)$, MCT was rapidly extended to single-particle observables such as $\phi_s(q,t)$ and the MSD.\cite{bengtzelius1984dynamics} The theory notably obtains a self-consistent approximate expression for the memory kernel $m_s(q,t)$ introduced in Eq.~\eqref{eq:GLE_tagged_ISF}. By analyzing its dimensional behavior, we can infer the MCT-predicted scaling of the MSD and MQD based on the general considerations presented above. 

\subsection{Large $d$ Scaling for the Hard Sphere Liquid}
\label{sec:MCT_scaling_HS}
In the case of simple hard spheres of diameter $\sigma$ in $d$ dimensions, MCT approximates the memory kernel of Eq.~\eqref{eq:GLE_tagged_ISF} as 
    \begin{equation}
        m_s^{\mathrm{HS}}(q,t) = \frac{\rho_0}{q^4}\int \frac{\mathrm{d}^dk}{(2\pi)^d}(\boldsymbol{q}\cdot\boldsymbol{k})^2S(k)c(k)^2\phi(k,t)\phi_{s}(|\boldsymbol{q}-\boldsymbol{k}|, t)
    \label{eq:ms_HS_MCT}
    \end{equation}
where $\rho_0$ denotes the number density, $c(k)$ the two-particle direct correlation function, and $S(k)$ the corresponding structure factor, which are related as $c(k) = \rho_0^{-1}[1-1/S(k)]$. In mode-coupling theories, the products of structural inputs in the integrand of Eq.~\eqref{eq:ms_HS_MCT} are commonly referred to as (static) vertices. We denote $\varphi \equiv \rho_0 V_d$ as the packing fraction, where $V_d$ is the volume of a $d$-dimensional sphere of unit diameter. (Without loss of generality, we henceforth set $\sigma=1$.) 
A low-$q$ expansion of Eq.~\eqref{eq:ms_HS_MCT} gives
    \begin{equation}
        m_0^{\mathrm{HS}}(t) = \frac{\rho_0}{(2\pi)^d}\frac{\Omega_{d}}{d}\int_0^{\infty} \mathrm{d}k k^{d+1}S(k)c(k)^2\phi(k,t)\phi_s(k,t)
    \label{eq:m0_MCT}
    \end{equation}
and 
    \begin{equation}
        m_2^{\mathrm{HS}}(t) = \frac{3\rho_0 \Omega_d}{(2\pi)^d d(d+2)} \int_0^{\infty}\mathrm{d}k k^{d+1} S(k)c(k)^2\phi(k,t)\mathcal{D}_d(k,t),
    \label{eq:m2_MCT}
    \end{equation}
where $\Omega_d$ is the surface of a $d$-dimensional sphere and 
    \begin{equation}
        \mathcal{D}_d(k,t) \equiv \frac{d-1}{3k} \pdv{ \phi_s(k,t)}{ k} + \pdv[2]{\phi_s(k,t)}{k}.
    \end{equation}
Note that static triplet correlations, which formally also contribute to $m_s^{\mathrm{HS}}(q,t)$, are here neglected because many-body correlations are exponentially suppressed with $d$.\cite{schmid2010,liu2021dynamics} Hence, MCT only requires the static structure factor as input to make quantitative predictions about the liquid dynamics.

In order to investigate the large-$d$ behavior of the equations above, it is convenient to introduce the dimensionally scaled momentum $k = %(\tilde{k}/\sigma) d = 
\tilde{k}d$. Because in the limit $d\rightarrow\infty$, the collective and self intermediate functions become equivalent, we may then replace $\phi(\tilde{k}d,t) \rightarrow\phi_s(\tilde{k}d,t)$, and $S(\tilde{k}d)\rightarrow1$. In the glass phase, i.e.\ the non-ergodic regime predicted by MCT, we further have that at long times $\phi_s(\tilde{k}d,t\rightarrow\infty)=\phi_s^{\infty}(\tilde{k}d)$. In Ref.~\citenum{schmid2010}, Schmid and Schilling have worked out that in these limits, MCT gives $\phi_s^{\infty}(\tilde{k}d) \propto \Theta(\tilde{k}_*-\tilde{k})$, where $\Theta(x)$ denotes the Heaviside function. As a result, the quantity $\tilde{k}_* = x\sqrt{d}$ provides a natural, dimensionally dependent ultra-violet (UV) cutoff for the theory. Numerical studies in large $d$ indicate that $x\approx 0.15$ is a reasonable choice.\cite{schmid2010, jin2015} All these results can be derived from the fact that the large $d$ limit of the direct correlation function, $c(r) = -\Theta(\sigma-r)$, can be shown to give
    \begin{equation}
        c(k) = - \left(\frac{2\pi}{k\sigma}\right)^{d/2}\sigma^d J_{d/2}(k\sigma) 
    \label{eq:asymptotic_structure}
    \end{equation}
in Fourier space, where $J_n(k\sigma)$ denotes the Bessel function of order $n$.\cite{frisch1999high} Using asymptotic expansions for the large order behavior of Bessel functions,\cite{abramowitz1965handbook, schmid2010} it is then possible to show that Eq.~\eqref{eq:m0_MCT} simplifies to
    \begin{equation}
        m_0^{\text{HS}}(t\rightarrow\infty) = d^{7/2} \times \frac{\tilde{\varphi}}{2\pi}\sqrt{4x^2 - \frac{1}{d}}
    \label{eq:UV_cutoff}
    \end{equation}
and analogously
    \begin{equation}
        m_2^{\text{HS}}(t\rightarrow\infty) = - \frac{2 d \tilde{\varphi}}{\pi} + \mathcal{O}(d^0),
    \label{eq:scaling_m2_glass_MCT}
    \end{equation}
where $\tilde{\varphi} = 2^d \varphi / d^2$ is the dimensionally rescaled packing fraction. Assuming that these scalings hold for finite times on either side of the transition, MCT then predicts that $m_0^{\text{HS}}(t)= d^{7/2} \times \tilde{m}_0(t)$ and $m_2^{\text{HS}}(t) = d \times \tilde{m}_2(t)$. The power-counting arguments presented in the previous section %Sec.~\ref{sec:general_NGP}
further lead to $\langle \Delta r^2(t) \rangle = d^{-5/2} \times \tilde{\Delta}_2(t)$  and $\langle \Delta r^4(t) \rangle = d^{-4} \times \tilde{\Delta}_4(t)$ with $\tilde{\Delta}_{2,4}(t)$ well behaved functions as $d\rightarrow\infty$. Furthermore, we find that MCT predicts that the NGP should scale as $\alpha_2(t) = d \times \tilde{\alpha}_2(t)$. Finally, within MCT we find that non-trivial physics emerges when the bare diffusion constant is of order $D_0 = d^{-7/2} \times \tilde{D}_0$, and that in this regime the long-time diffusion constant should go as $D = d^{-7/2}\times \tilde{D}$ where $\tilde{D}_0$ and $\tilde{D}$ are finite in the large dimensional limit.\cite{footnote2} \\

Overall, from the above MCT analysis, we find that as $d\rightarrow\infty$, MCT predicts dimensional scalings that disagree with the exact DMFT results. 
In particular, the dynamics predicted by MCT is \textit{strongly non-Gaussian}, $\alpha_2(t) = d \times \tilde{\alpha}_2(t)$, whereas it is expected to become Gaussian.

\subsection{Large $d$ Scaling for the Mari--Kurchan Liquid}

As mentioned in the introduction, MCT can also be used to study other structured finite-$d$ fluid glass formers. A particularly interesting model was proposed by Mari and Kurchan to interpolate continuously between a minimally structured and a realistic HS fluid in finite $d$ by considering  particles with pairwise, randomly shifted interactions.\cite{mari2011dynamical}  More specifically, the MK interaction potential between two particles located at $\boldsymbol{r}_i$ and $\boldsymbol{r}_j$ takes the form $U(|\boldsymbol{r}_i-\boldsymbol{r}_j - \bm{\Lambda}_{ij}|)$
with $\bm{\Lambda}_{ij}$ a uniformly distributed random vector with variance $\sigma_{\Lambda}$.\cite{mari2011dynamical} Note that the limit of vanishing variance, $\sigma_{\Lambda}\rightarrow0$, recovers standard hard spheres, whereas the limit of infinite variance results in a minimally structured fluid, akin in that sense to hard spheres in the limit $d\rightarrow\infty$.\cite{mari2011dynamical} While no exact solution for the structure nor the dynamics of this model is known for arbitrary $\sigma_{\Lambda}$, increasing the variance is expected to reduce the  fluid structure monotonically. The MK model therefore offers the possibility to isolate the role played by local structure on the dynamics of finite-$d$ systems. Furthermore, in the limit $d\rightarrow\infty$, this model is known -- for all $\sigma_{\Lambda}$ -- to have precisely the same description as hard spheres.

In order to analyze MCT for the MK model, the theory needs to be modified slightly to account for the absence of many-body structural and dynamical correlations when $\sigma_{\Lambda}$ is of the order of the system size. In this limit, the resulting mode-coupling scheme approximates the memory kernel as (see Ref.~\citenum{laudicina2024MK} for details): 
    \begin{equation}
        m_s^{\mathrm{MK}}(q,t) = \frac{\rho_0}{q^4}\int \frac{\mathrm{d}^dk}{(2\pi)^d}(\boldsymbol{q}\cdot\boldsymbol{k})^2c(k)^2\phi_{s}(k,t)\phi_{s}(|\boldsymbol{q}-\boldsymbol{k}|, t).
    \label{eq:MK_memory_kernel}
    \end{equation}
Note the similarity between this last expression and Eq.~\eqref{eq:ms_HS_MCT} with the intermediate scattering function $\phi \rightarrow \phi_s$ replaced and the structure factor $S(k)$ set to unity in the vertices. The direct correlation function $c(k)$ is here given by Eq.~\eqref{eq:asymptotic_structure}. In the limit $d\rightarrow\infty$, the microscopic MCT equation of motion for hard spheres given in Eq.~\eqref{eq:UV_cutoff} is hence recovered. Within a mode-coupling scheme, the MK model therefore also belongs to the same universality class as hard spheres.\cite{schmid2010} This equivalence, which is also expected from static considerations,\cite{mari2011dynamical} implies that the various MCT dimensional scaling analyses previously considered carry over to the MK model. 

\subsection{Numerical Confirmation of the Hard-Sphere Universality Class Captured by MCT}

\begin{figure}
    \centering
    \includegraphics[width=0.7\columnwidth]{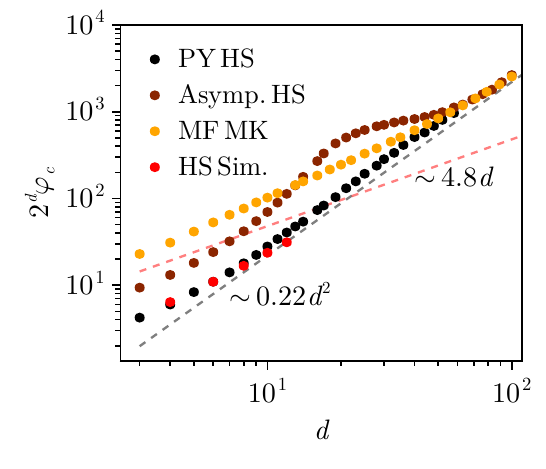}
    \caption{\textit{Critical MCT packing fraction $\varphi_c(d)$ for various model glass formers in the hard-sphere universality class: hard spheres with Percus-Yevick structure (black points);  hard spheres with the asymptotic (large-$d$) structure from Eq.~\eqref{eq:asymptotic_structure} (brown points); the Mari--Kurchan model (orange points); asymptotic high $d$ scaling (dashed black line). Hard sphere simulation results (red points) and the exact asymptotic result from DMFT (dashed red line) are provided as reference.}}
    \label{fig:MCT_crit_pt_scaling}
\end{figure}

Next, we numerically validate the existence of the HS universality class by studying the $d$-dependent behavior of the MCT-predicted critical packing fraction of the ergodic--nonergodic transition, $\varphi_c(d)$. Let us first recall that in the case of high-dimensional hard spheres, two previous MCT studies had focused on the numerical solutions of the MCT equations with the large-$d$ asymptotic structure of a liquid combined with either the microscopic MCT equations\cite{ikeda2010} or the large-$d$ asymptotic MCT equations,\cite{schmid2010} both yielding
\begin{equation}
    \varphi_c(d)\propto 2^{-d}d^2.
    \label{eq:MCT_crit_pt_scaling}
\end{equation}
We have reproduced these results in Fig.~\ref{fig:MCT_crit_pt_scaling}, showing that convergence to the analytical scaling for $\varphi_c(d)$ is reached for $d\gtrsim100$.  Here, we complement these prior studies and also solve for the microscopic MCT with the explicitly $d$-dependent Percus-Yevick (PY) structure for hard spheres, up to $d=60$ (see Appendix for numerical details). Interestingly, as shown in Fig.~\ref{fig:MCT_crit_pt_scaling}, we find that this fully microscopic calculation  approximately follows the analytical and asymptotically valid (for $d\rightarrow \infty$) scaling of $\varphi_c(d)$ at all dimensions considered. 
We have verified that using the hypernetted-chain approximation of the liquid structure in MCT gives similar results. We attribute the difference with the results of Refs.~\citenum{ikeda2010, schmid2010} to the use of different structural inputs to the mode-coupling equations, which in these cases followed from the asymptotic expression Eq.~\eqref{eq:asymptotic_structure}. It must be noted, however, that even in $d=60$ the PY structure differs markedly from the asymptotic one, and the agreement between our MCT PY HS results and the asymptotic scaling across all dimensions might (at least partly) bear a fortuitous origin. 

Additionally, we demonstrate that the MK model, treated within the mode-coupling approximation, also converges to the required analytic asymptotic behavior for large $d$. We show the numerical results obtained by using Eq.~\eqref{eq:asymptotic_structure} as input for the MK MCT in Fig.~\ref{fig:MCT_crit_pt_scaling}. We note that convergence to the analytical scaling is reached at around $d\approx 100$, in accordance with HS MCT using the same asymptotic structure.\cite{schmid2010, ikeda2010}
Our results also show that the critical packing fraction of the MK model is systematically larger than that of HS systems across all dimensions, until convergence to the analytical scaling is attained. We can understand this difference with respect to structured fluids by recognizing that the mode-coupling vertices are density independent for the MK model, unlike for HS. In the latter, there is an intricate relation between structural features and bulk density in the mode-coupling vertices which ultimately drives the solidification into an amorphous state. For minimally structured fluids like the MK model however, the density independent vertices imply that dynamical arrest is instead purely driven by the bulk density prefactor, as seen in Eq.~\eqref{eq:MK_memory_kernel}.

Although the asymptotic large-$d$ behavior of MCT disagrees with the exact results from the DMFT analysis,\cite{liu2021dynamics,biroli2022} which instead predicts that $\varphi_c(d) \propto 2^{-d} d$, the MCT description is nevertheless internally self-consistent. In particular, both MCT\cite{schmid2010, ikeda2010, jin2015} and DMFT\cite{mari2011dynamical,biroli2021,biroli2022} find that simple hard spheres, the MK model, and also the random Lorentz gas all belong to the same (dynamical) universality class in the limit $d\rightarrow\infty$. 
Identifying the origin of the discrepancy between the scaling behaviors of $\varphi_c(d)$ within MCT and DMFT would be a first step towards bringing consistency to the description of all three systems at once.

\subsection{Recovering the Appropriate Dimensional Scalings within MCT}
The source of the erroneous MCT prediction for the dimensional scaling of $\varphi_c(d)$ can be traced back to the natural $d$-dependent UV cutoff seemingly embedded within the theory, as exemplified in Eq.~\eqref{eq:UV_cutoff}, for which the momentum integrals run up to a finite value in finite $d$. Previous work by Ikeda \& Miyazaki\cite{ikeda2010} has demonstrated that the erroneous dimensional scaling of $\varphi_c(d)$ can be remedied by enforcing a Gaussian ansatz for the NEP. Inspired by their approach, we next show that such a Gaussian ansatz can also recover the appropriate $d$-dependent scaling for both the MSD and the MQD (and their associated memory kernels) within MCT. Explicitly, let us denote as $\overline{\cdot\cdot}$ quantities computed with a Gaussian ansatz. From Eq.~\eqref{eq:ms_HS_MCT}, it is possible to show that 
    \begin{equation}
        \overline{m}_0^{\text{HS}} = \frac{\bar{\varphi}d^2}{2\pi} \times \sqrt{\frac{\pi}{\bar{R}}}e^{-\bar{R}/4},
        \label{eq:m0HS}
    \end{equation}
where we have used that $R$ scales such that $R = d^{-1} \bar{R}$ is finite in the limit $d\rightarrow\infty$ and $\bar{\varphi} \equiv 2^d \varphi / d$. The latter scaling can be self-consistently verified.\cite{ikeda2010,footnote2} We therefore obtain that $\overline{m}_0^{\text{HS}}(t) = d^{\upsilon}\times\overline{m}_0(t)$ with $\upsilon=2$. An analogous calculation gives $\overline{m}_2^{\text{HS}}(t) = d^{\lambda}\times \overline{m}_2(t)$ with $\lambda = 0$ and where $\overline{m}_2(t)$ is positive, unlike its microscopic MCT analogue above [see Eq.~\eqref{eq:scaling_m2_glass_MCT}]. The $d$-dependent scalings then agree with the DMFT predictions, as summarized in Table~\ref{tab:scaling_summary}. We note, however, that the Gaussian ansatz is known to give the wrong numerical prefactor for the $d$-dependent scaling of $\varphi_c(d)$.\cite{ikeda2010} Because we anticipate the same to be true for the other quantities considered in this work, exact equivalence with the DMFT is not to be expected. In particular, this means that we are unable to explicitly verify if the appropriate dimensional scaling of the NGP is recovered, which as mentioned above  depends on the vanishing of the pre-factor of the leading order term. 
{\renewcommand{\arraystretch}{1.2}
\begin{table}
    \centering
\caption{Summary of asymptotic large-$d$ scalings of various quantities for DMFT and microscopic MCT  as well as MCT with a Gaussian ansatz for the hard-sphere universality class. Note that the MCT predictions stem from power-counting arguments and are therefore strict upper bounds.}
    \begin{tabular}{|c|c|c|c|}
    \hline
         Quantity & DMFT (exact) & MCT & MCT Gaussian \\ \hline
         $\langle \Delta r^2(t)\rangle$ &  $d^{-1} \times \hat{\Delta}_2(t)$ & $d^{-5/2} \times \tilde{\Delta}_2(t)$ & $d^{-1} \times \overline{\Delta}_2(t)$\\
         $\langle \Delta r^4(t)\rangle$ & $d^{-2} \times \hat{\Delta}_4(t)$ & $d^{-4}\times \tilde{\Delta}_4(t)$ & $d^{-2}\times \overline{\Delta}_4(t)$\\
        $D$ & $d^{-2}\times \hat{D}$ & $d^{-7/2}\times \tilde{D}$ & $d^{-2}\times \overline{D} $\\
         $\alpha_2(t)$ & $d^{-1} \times \hat{\alpha}_2(t)$  & $d \times \tilde{\alpha}_2(t)$ & $d^{0}\times \overline{\alpha}_2(t)$\\
         $m_0(t)$ & $d^{2} \times \hat{m}_0(t)$ & $d^{7/2}\times \tilde{m}_0(t)$ & $d^{2}\times \overline{m}_0(t)$\\   
         $m_2(t)$ & $d^0\times \hat{m}_2(t)$ & $d \times \tilde{m}_0(t)$ & $d^0\times \overline{m}_2(t)$\\
         \hline
    \end{tabular}
\label{tab:scaling_summary}
\end{table}
}

\section{Numerical Simulations}
\label{sec:numerics}
We next seek to compare the MCT predictions for high-dimensional hard spheres and the MK model, in particular for the MSD and the NGP, with the dynamics obtained from particle-resolved computer simulations. For the MK model, we re-use the data published in
Ref.~\citenum{charbonneau2024dynamics}. Numerical simulation results for hard spheres in various spatial dimensions are obtained using the event-driven molecular dynamics package described in Ref.~\citenum{charbonneau2024dynamics} supplemented with an implementation of $D_d$ (or checkerboard lattice) periodic boundary conditions~\cite{charbonneau2022dimensional}. (Recall that $D_3$ is equivalent to a face-centered cubic lattice and $D_4$ is the densest sphere packing in $d=4$.) As described in Ref.~\citenum{charbonneau2022dimensional}, this scheme minimizes finite-size corrections by simulating box shapes that correspond to the Wigner--Seitz cell of checkerboard packings in various $d$. As a result, numerical simulations of hard spheres can be run up to $d=12$ with system sizes up to $N=8000$, with only minimal finite-size corrections.~\cite{charbonneau2013} 

In general, initial (equilibrated) configurations are obtained from the configurations obtained for Ref.~\citenum{charbonneau2024dynamics}, and initial velocities are randomly assigned from the Maxwell--Boltzmann distribution. The system sizes used here therefore match those used in that prior work. More specifically, simulations at volume fraction $\varphi$ are started from the nearest reference configuration with $\varphi_0 \ge \varphi$, and the sphere radius is instantaneously shrunk to achieve the target density. For configurations at volume fractions higher than those explored by Ref.~\citenum{charbonneau2024dynamics}, simulations start from the densest available configuration, and then employ the collision-driven compression scheme of Ref.~\citenum{skoge2006packing} with a sphere radius expansion rate of $\alpha = 0.001$, i.e. sphere radius $r(t) = (1 + \alpha \sigma t) r_0$ where $\sigma$ sets the unit of length in simulation, until the target density is reached. In both cases, systems are then equilibrated  such that either $d \times \langle \Delta r^2(t)\rangle$ reaches $10^1$ or $1/4$ of the total run time, whichever comes first before collating results. Observables are then extracted from the remaining trajectory and averaged over at least 20 independent runs. Trajectories range in duration from $10^5$ in $d=4$ to $10^4$ in $d=12$.

\section{Behavior of the MSD and the NGP: Results from MCT \& Numerical Simulations}
\label{sec:results}
In this section, results from numerical simulations of hard spheres are compared with MCT predictions at finite times, first qualitatively %in %Sec.~\ref{sec:qualitative}, 
and then more quantitatively by considering the (dynamical) critical exponents and other expected scaling laws.% in Sec.~\ref{sec:criticalexp} . 

\begin{figure*}
    \includegraphics[width=\textwidth]{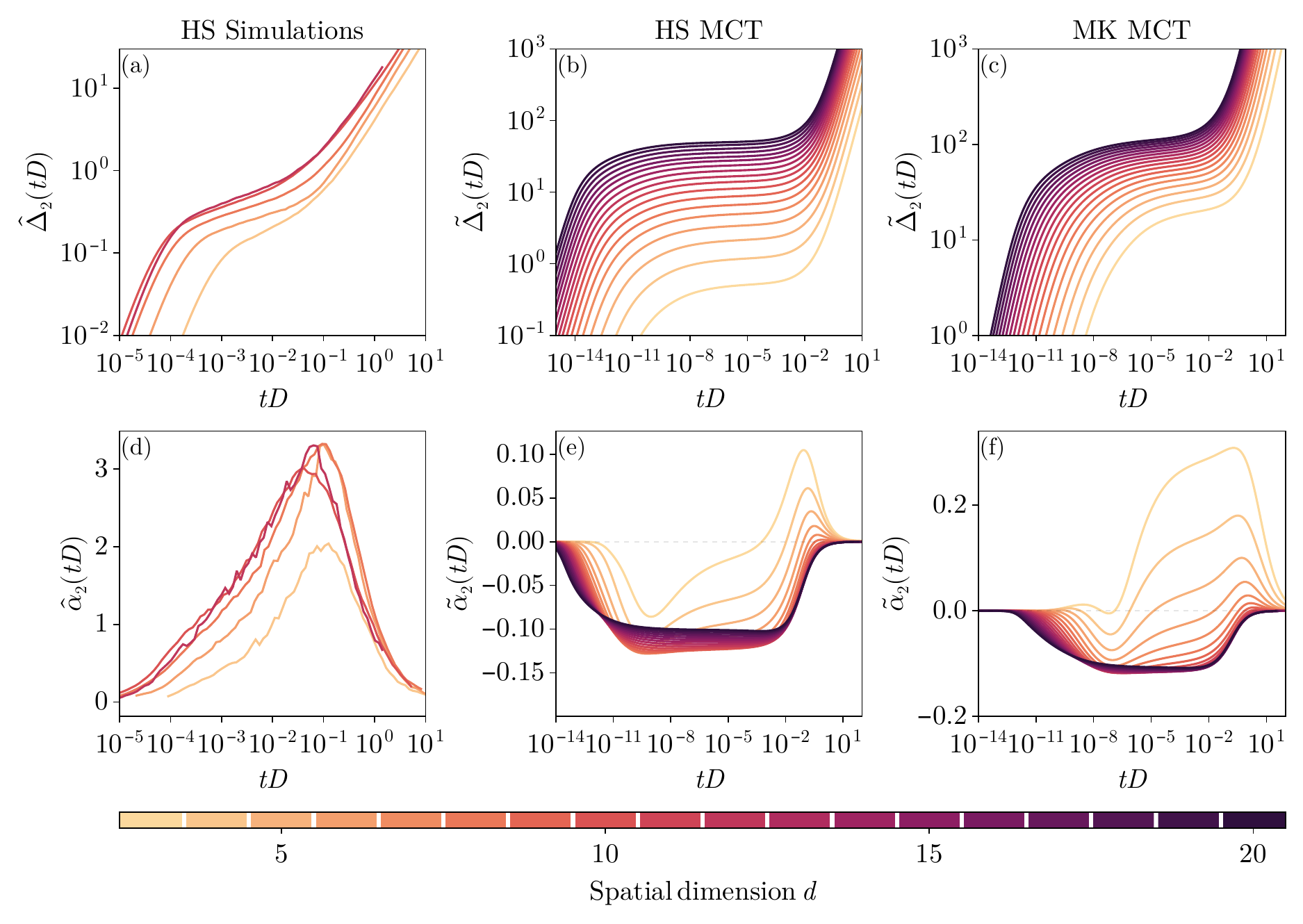}
    \caption{Dimensionally scaled MSD (a)--(c) and NGP (d)--(f) for simulations of hard-spheres, MCT of hard-spheres and MCT of the mean-field MK model. MCT results are for $d=3$ to 20 at fixed relative distance $\varepsilon=10.0^{-5}$ to the critical point;  
    simulation results are for $\varepsilon \sim \mathcal{O}(10^{-2})$ in $d=4,6, \ldots, 12$.
    Although finite $d$ corrections to MSD scaling can be noted, the asymptotic limit is nearly attained in both cases. Note that the time axes have been rescaled by the respective values of the long-time diffusion constant $D$.}
\label{fig:qualitative_comparison_MCT_DMFT}
\end{figure*}

\subsection{Qualitative Comparison}
\label{sec:qualitative}
Figure~\ref{fig:qualitative_comparison_MCT_DMFT} (a)-(b) shows dimensionally rescaled MSDs obtained both from MCT [$\tilde{\Delta}_2(t) \equiv d^{-5/2} \langle \Delta r^2(t)\rangle$] and from HS simulations [$\hat{\Delta}_2(t) \equiv d^{-1} \langle \Delta r^2(t)\rangle$]. We can see that the MCT predictions are in qualitative agreement with the MSDs from HS simulations. In particular, sufficiently close to the critical point, in all $d$ the short-time (ballistic) regime gives way to a plateau (caging) regime, before diffusion finally takes over at long times. As the spatial dimension increases, we expect convergence to the respective master curve $\hat{\Delta}_2(t)$ for the simulations and $\tilde{\Delta}_2(t)$ for MCT.
Although finite-$d$ corrections are visible -- for MCT, the asymptotic behavior is expected to set in for $d\gtrsim 100$ --  the trend cleanly tends towards convergence. A similar dimensional collapse is noted for computer simulation results in Fig.~\ref{fig:qualitative_comparison_MCT_DMFT}-(a), with results for $d=10$ and 12 nearly collapsing already. 

For the NGP, however, plotted in Fig.~\ref{fig:qualitative_comparison_MCT_DMFT} (d)-(e), the agreement between theory and simulations is more equivocal. As expected, numerical simulation results [panel (d)] peak at times that roughly correspond to the end of the caging regime and vanish at both short and long times. For a fixed $\varepsilon$, the peak of the rescaled NGP steadily increases and then saturates as $d$ increases 
-- consistent with previous reports,\cite{adhikari2021} and the $\hat{\alpha}_2(t)$ curves collapse for $d\geq8$. Note that saturation of dimensionally rescaled quantities corresponds to unscaled quantities decreasing. Thus, one finds that the dynamics of the fluid become increasingly more Gaussian as $d$ increases, consistent with DMFT expectations predicting a vanishing NGP in the limit $d\rightarrow\infty$.\cite{liu2021dynamics} In moderately high dimensions, $d\leq 8$, MCT [panel (e)] predicts that the NGP also peaks at times comparable to the relaxation time and vanishes at both short and long times, and at fixed $\varepsilon$, the peak height also steadily decreases as $d$ increases. That apparent consistency, however, disregards the negative dip of the NGP from MCT at short times. When first noted, this negative dip was attributed to the inability of MCT to appropriately account for short-time correlations between particle collisions, which should in principle be encoded in an additional, wave-vector dependent prefactor to the friction term in Eq.~\eqref{eq:GLE_tagged_ISF}.\cite{fuchs1998asymptotic} This interpretation can be tested by considering MCT results for the minimally structured MK model, for which such correlations should be strongly suppressed. As shown in Fig.~\ref{fig:qualitative_comparison_MCT_DMFT}-(f), the NGP for the MK model predicted by MCT does indeed remain (nearly) positive at all times in $d=3$. This behavior is analogous to that previously reported for the NGP of a small tagged particle immersed in a fluid of larger particles, and thus further supports the friction-based explanation.\cite{fuchs1998asymptotic} The dimensional trend, however, does not support that hypothesis. Already in $d=4$, the NGP for the MK model presents a short-time negative dip, and the effect grows with $d$ for both HS and the MK model, even though structural correlations in the former then steadily decrease and remain immaterial in the latter. At present, the precise origin of this initial negative dip within the MCT predictions remains unclear. 

In the vicinity of the critical regime, MCT also predicts that the NGP further grows in a two-step fashion,\cite{fuchs1998asymptotic} thus echoing the two power-law regimes that describe the approach and the departure from the MSD plateau, respectively. 
Because DMFT shares the same critical description as MCT -- albeit with different exponents\cite{maimbourg2016solution} -- one should then expect a qualitatively similar behavior to emerge, but quantitative results have yet to be obtained for DMFT. Nonetheless, the pre-peak growth of the $d=4$ and $d=6$ simulation results in Fig.~\ref{fig:qualitative_comparison_MCT_DMFT}-(d) is consistent with this expectation. The limited extent of the caging regime accessible in simulations -- especially as $d$ grows -- prevents a more definitive identification of the two scaling regimes. Over this dimensional range, the general trend from MCT is therefore consistent with the simulation results.

For higher dimensions, $d\geq 8$, MCT predictions are clearly qualitatively incorrect. The NGP is then negative for nearly all times, except at short and long times, when it vanishes. While it has been argued that $d=8$ should be the upper-critical dimension of MCT,\cite{Biroli2007,Franz2012,berthier2020} we believe this to be a mere coincidence with the present observations. This discrepancy likely follows from two separate features of MCT. First, MCT predicts a small transient and unphysical negative behavior for the MQD. This pathological prediction robustly appears for both HS and the MK model at least up to $d=20$ (see Appendix for an additional discussion). Although MCT is also known to predict a similarly unphysical negative dip in the self part of the Van Hove distribution function,\cite{ikeda2010} the precise nature of this breakdown remains unknown. Second, one can show that the NGP, close to the dynamical transition, takes the following asymptotic form in the ergodic phase\cite{fuchs1998asymptotic}: 
    \begin{equation}
    \begin{split}        
        \alpha_2(t) =&\ \alpha_2^{\mathrm{c}} + h_{\mathrm{NGP}}G(t) + h_{\mathrm{NGP}}\left(H(t) + K_{\mathrm{NGP}}G(t)^2 - |\varepsilon|  \hat{K}_{\mathrm{NGP}}\right) 
    \end{split}
    \label{eq:NGP_asymptotic}
    \end{equation}
where $G(t),\ H(t)$ are leading and next-to-leading order scaling functions (in $\varepsilon$). We recognize that such an expansion is valid in arbitrary $d$ since the MCT scenario is unaffected by dimension.\cite{schmid2010} The scaling functions describe the power-law approach and departure from the plateau of the MSD, respectively. The quantities $h_{\mathrm{NGP}}$, $K_{\mathrm{NGP}}$ and $\hat{K}_{\mathrm{NGP}}$ are known as critical amplitudes and $\alpha_2^{\mathrm{c}}$ is the long-time limit of the NGP evaluated at the critical point. Note that these quantities can all be computed from structural inputs and are also explicitly dependent on $d$.
Here, we find that as $d$ increases, the time window in which the second term contributes seems to vanish, leading us to conclude that $h_{\mathrm{NGP}}$ must vanish (since $G(t)$ does not, see Appendix for additional details). As a consequence, the growth of the NGP is completely hampered in the caging regime within MCT. A numerical verification of this hypothesis would require an explicit computation of the critical amplitudes, which is beyond the scope of the current work. In addition, we have found that the long-time value of the NGP from MCT in the non-ergodic phase is systematically negative for all $d$, generalizing the results of Ref.~\citenum{fuchs1998asymptotic} for $d=3$, whereas a positive value is observed in simulations.

These observations lead us to conclude that for both HS systems and the MK model, the NGP predicted by MCT exhibits unphysical features in all $d$. These shortcomings grow increasingly significant as $d$ increases. While not studied here specifically, it is expected that similar shortcomings would be observed in the MCT of other systems that fall into the same hard sphere universality class (such as the random Lorentz gas). 

\subsection{Diverging Timescale of Critical Fluctuations}
\label{sec:criticalexp}

Empirically, a growing timescale $t_{\mathrm{peak}}$ can be associated with the peak of the NGP as the density is increased. This timescale has also been shown to be a good estimate of the structural relaxation time\cite{flenner2005relaxation2, mizuno2010lifetime}, measured here via the magnitude of the long-time diffusion constant $D$. In theories such as MCT and DMFT,  the vanishing of the diffusivity near the dynamical arrest is known to be controlled by a critical exponent $\gamma$, such that $D \propto |\varphi - \varphi_c|^{\gamma}$. For this reason, one also naively expects that $t_{\mathrm{peak}} \propto |\varphi - \varphi_c|^{-\gamma}$. 
Within MCT, $\gamma$ is material (and $d$) dependent because its value is controlled by the fluid structure; a similar material (and $d$) dependence is expected from DMFT, but has yet to be specifically considered. More importantly, although MCT predicts a critical dynamical arrest in all $d$, recall that the critical point is avoided in finite $d$ and thence becomes a dynamical crossover that leaves a pseudo-critical regime in a portion of parameter space.\cite{berthier2011} 
In this regime, the exponent $\gamma$ should govern both the timescale associated with the critical dynamical fluctuations and the vanishing of the diffusivity. Consequently, one can estimate $\gamma$ in two different but supposedly equivalent ways in both mode-coupling studies and computer simulations.

\begin{figure}
    \centering
    \includegraphics[width=0.75\linewidth]{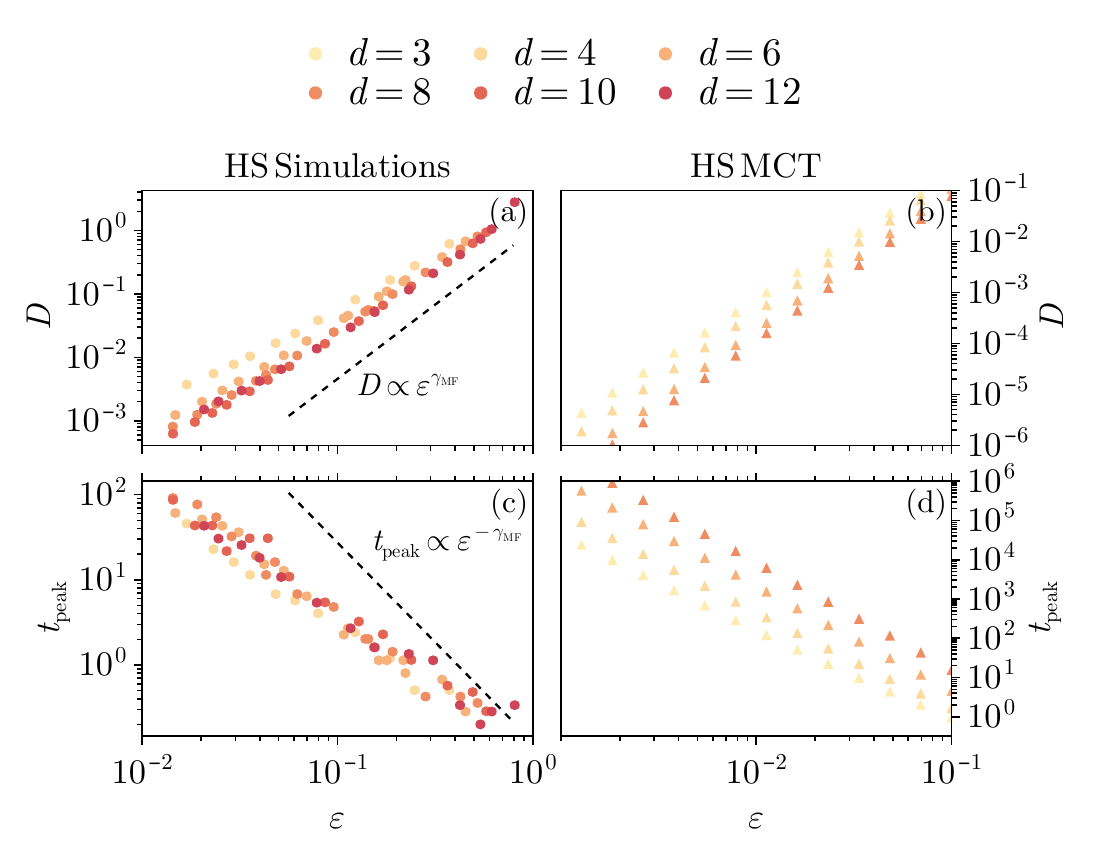}
    \caption{Vanishing of the diffusivity $D$ as measured from (a) HS simulations and  (b) MCT. Growing timescale associated with single particle critical fluctuations $t_{\mathrm{peak}}$ as measured from (c) HS simulations and  (d) MCT. All panels present observables as a function of the relative distance to the critical point $\varepsilon\equiv |\varphi-\varphi_c|/\varphi_c$. Dashed lines indicate the exact predictions of DMFT in the limit $d\rightarrow\infty$.}
    \label{fig:t_peak_MCT_sims}
\end{figure}

We first qualitatively compare the dimensional dependence of $\gamma$ calculated from MCT with simulation results for HS and the MK model. Figure~\ref{fig:t_peak_MCT_sims} demonstrates that both $D$ and $t_{\mathrm{peak}}$ indeed show a transient power-law like regime in simulations of HS [left column, panels (a)-(c)] and that a similar behavior is observed for our mode-coupling calculations [right column, panels (b)-(d)]. We also find that results from HS simulations appear to converge towards the expected $d\rightarrow\infty$ behavior. In all cases, we can extract the value of the exponent $\gamma$ by fitting $D$ and $t_{\mathrm{peak}}$ close to $\varphi_c$. For HS simulations, $\varphi_c$ is also undetermined and hence a two-parameter fit is normally employed, but we here instead use the values of $\varphi_c$ reported in Ref.~\citenum{charbonneau2022dimensional}. For the MK model, $\varphi_c$ is known to vary little from its value in the limit $d\rightarrow\infty$~\cite{charbonneau2014hopping}, and can therefore be estimated as in Ref.~\citenum{charbonneau2022dimensional}. While this fitting procedure is nearly exact for MCT calculations\cite{footnote3} (limited only by the numerical precision used to solve its equations), it is more technically challenging for low $d$ simulations given that both perturbative and non-perturbative (activated) events are significant near the avoided critical point.\cite{charbonneau2014hopping, charbonneau2022dimensional} We also note that $t_{\mathrm{peak}}$ (and the corresponding $\gamma$) can only be measured from MCT in $d\leq 8$, as the numerical solutions no longer peak beyond this point (as per the discussion above). 

\begin{figure}[ht]
    \centering
    \includegraphics[width=0.75\columnwidth]{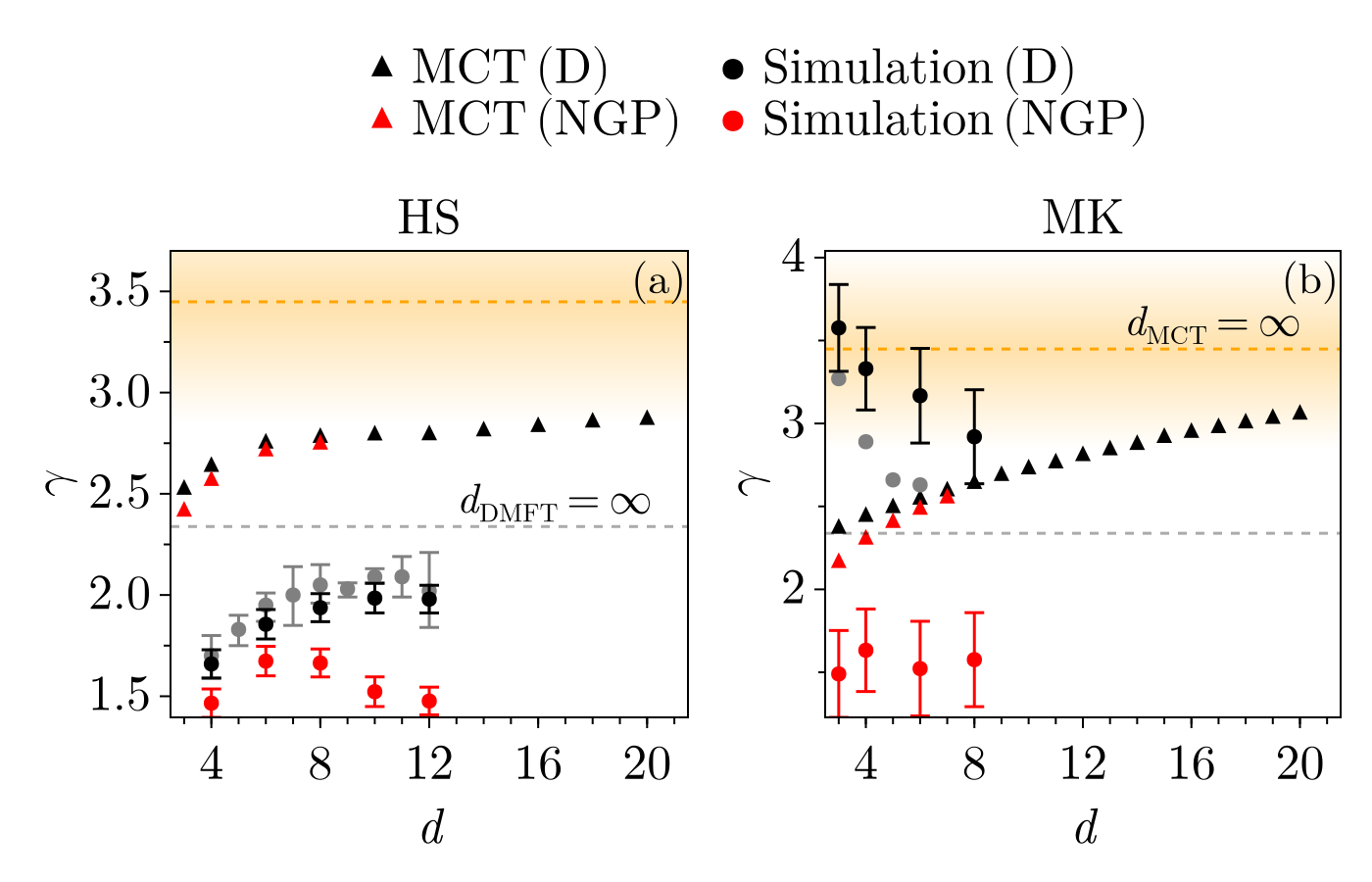}
    \caption{(a)-(b) Critical dynamical exponent $\gamma$ governing the decay of the diffusivity, $D\propto |\varphi - \varphi_c|^{-\gamma}$, (black) and the divergence of the NGP peak time, $t_{\mathrm{peak}}\propto |\varphi - \varphi_c|^{-\gamma}$, (red) for HS and the MK model from MCT (triangles) and computer simulations (circles).  In all panels, the shaded region provides an estimate of the MCT prediction in the limit $d\rightarrow\infty$ and  the horizontal dashed line indicates the exact DMFT result in that same limit.  Earlier results from Ref.~\citenum{charbonneau2022dimensional} for HS and Ref.~\citenum{charbonneau2014hopping} for the MK model (gray) are included. (For the MK model, error bars  were not reported, but they are presumed to be large.) Note that the error bars capture only the standard error of the fit. The true error is larger if one factors in the determination of $\varphi_c$ in finite $d$.}
\label{fig:dynamical_exponent_gamma_HS_MK}
\end{figure}

Despite these technical caveats, robust conclusions can be drawn from comparing theory with numerics. The results are shown in Fig.~\ref{fig:dynamical_exponent_gamma_HS_MK}. First, we find that MCT predicts essentially equivalent exponents for both methods used to measure it, as shown by the red and black triangles in Fig.~\ref{fig:dynamical_exponent_gamma_HS_MK}. Second, both MCT and HS simulation results show that $\gamma$ (irrespective of how it is measured) monotonically increases with $d$ and asymptotes to its respective $d\rightarrow\infty$ value. (The numerical results above are also consistent with earlier estimates for $\gamma$.\cite{charbonneau2022dimensional}.) We also note that measurements from $t_{\mathrm{peak}}$ in simulations exhibit a noisier convergence towards the asymptotic value [see red points in Fig.~\ref{fig:dynamical_exponent_gamma_HS_MK}-(a)].  
The asymptotic $d\rightarrow\infty$ values of the exponent, however, are significantly different: $\gamma_{\mathrm{MCT}}(d\rightarrow\infty) \approx 3.45$ -- as estimated from the critical parameter\cite{schmid2010} $0.8 < \lambda_{\mathrm{MCT}}(d\rightarrow\infty) < 0.9$ -- while the exact value from DMFT is $\gamma_{\mathrm{MF}} \approx 2.34$.\cite{maimbourg2016solution} Given the markedly distinct dimensional scaling given by the two theories (see Table~\ref{tab:scaling_summary}), this numerical discrepancy is not particularly surprising and even somewhat encouraging in that the two predictions are of the same order of magnitude. 

The results for the MK model -- shown in Fig.~\ref{fig:dynamical_exponent_gamma_HS_MK}-(b) -- are more problematic. Although monotonic convergence to the $d\rightarrow\infty$ values is preserved, the qualitative trend between MCT and simulation results no longer matches. While MCT predictions increase with $d$, simulation estimates from the diffusivity decrease, in agreement with earlier works.\cite{charbonneau2014hopping} Interestingly, when measured in simulations from $t_{\mathrm{peak}}$, we find that $\gamma$ appears to approach its asymptotic value from below, albeit very slowly if at all. 

While the various estimates of $\gamma$ would be (na\"ively) expected to coincide for both systems, the effect could here be masked by a combination of pre-asymptotic corrections to the pseudo-critical scaling, numerical errors in $\varphi_c$, as well as activated processes which may impact the diffusivity and non-Gaussian behavior in different ways. Given that both HS and the MK model present a similar underestimation of $\gamma$ when measured from critical fluctuations, the first and third explanations are more likely than the second, but (even) higher $d$ simulations would be needed to resolve these effects more clearly. We also cannot exclude that other physical effects might also play a role.

\subsection{Magnitude of Critical Fluctuations}
The growth of the NGP peak height in the (pseudo-)critical regime also provides some physical insight. First, recall that the NGP is fundamentally a single-particle observable, and that prior works by some of us\cite{biroli2022,charbonneau2024dynamics} have demonstrated that the NGP can be interpreted as the correlation of particle displacements across different spatial directions (or components). In systems dominated by single-particle dynamics (such as HS and the MK model as $d$ increases), once displacements along different components become fully correlated, the NGP peak $\alpha_2^*(\varphi ; d)$ should saturate to a constant $\alpha_2^{\mathrm{sat.}}(d)$ which in turn diverges linearly with $d$, i.e., with the number of such components. 
In the limit $d\rightarrow\infty$, we then expect $\alpha_2^*(\varphi ; d\rightarrow \infty) \propto |\varphi-\varphi_c|^{-1}$.\cite{biroli2022} 

In the presence of collective effects such as facilitation or multi-particle dynamical heterogeneity, however, the NGP can exceed predictions based off of the single-particle description. For instance, the NGP of the MK model, which permits facilitation,\cite{charbonneau2014hopping,nishikawa2022collective} does not plateau as cleanly as that of the (purely single-particle in all $d$) random Lorentz gas.\cite{charbonneau2024dynamics} Given the more pronounced contributions of collective effects in HS than in the MK model, an even greater excess NGP would be expected, although quantifying this excess is challenging given the (still) unclear quantitative mapping of this quantity between the random Lorentz gas and multi-particle systems.~\cite{charbonneau2024dynamics} 

\begin{figure}
    \centering
    \includegraphics[width=0.75\columnwidth]{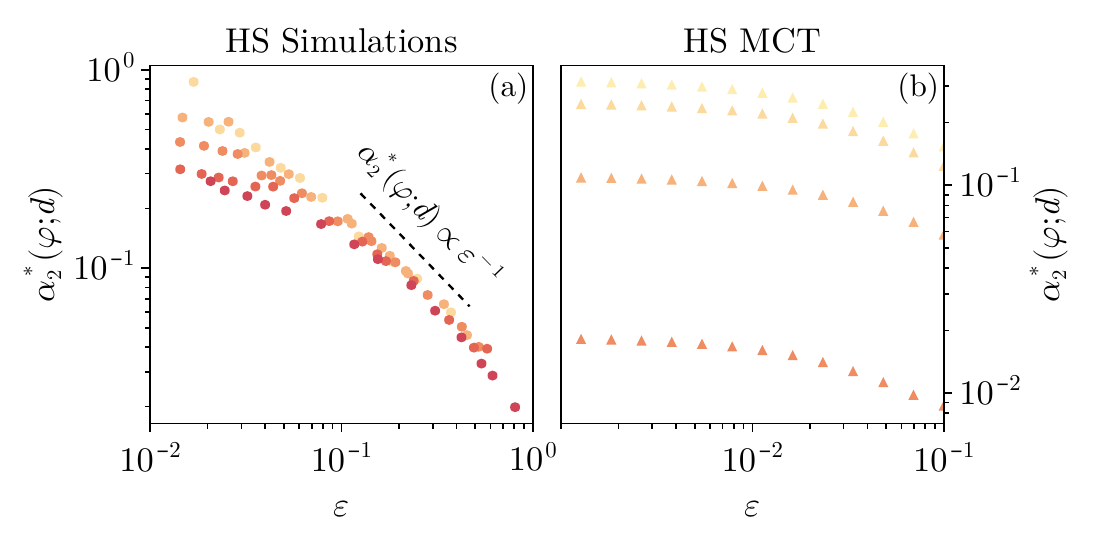}
    \caption{Scaling behavior of the peak of the NGP for HS in $4\leq d\leq12$ as a function of the relative distance to the dynamical critical point $\varepsilon \equiv |\varphi-\varphi_c|/\varphi_c$ as (a) measured in  numerical simulations and (b) predicted by MCT. The expected $d\rightarrow\infty$ scaling with critical exponent $\alpha_2^*(\varphi ; d) \propto \varepsilon^{-1}$ is provided as reference (dashed line).}
\label{fig:NGP_peak_scaling_MCT_sim}
\end{figure}

\begin{figure}
    \centering
    \includegraphics[width=0.5\columnwidth]{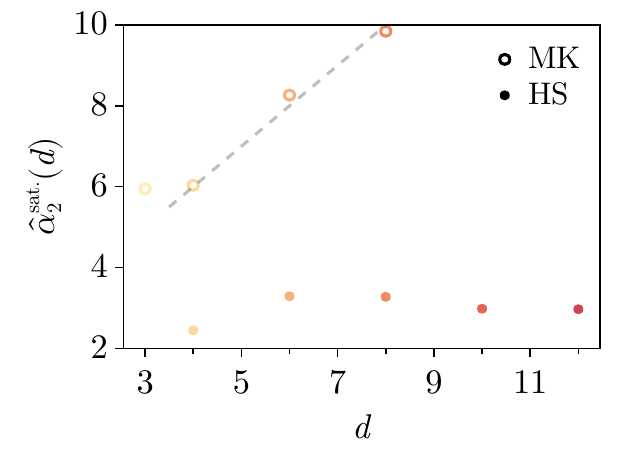}
    \caption{The dimensionally scaled saturation height of the NGP $\hat{\alpha}_2^{\mathrm{sat.}}(d)$ for the simulated HS systems and MK models. Data for the MK model are reproduced from Ref.~\citenum{charbonneau2024dynamics}. The saturation height is estimated from averaging over its value for the three smallest $\varepsilon$ values available. Although the linear relation $\hat{\alpha}^{\mathrm{sat.}} \propto d$ is clearly visible for the MK model (see dashed line), it is not so for HS.}
    \label{fig:NGP_sat_HS_MK}
\end{figure}

Figure~\ref{fig:NGP_peak_scaling_MCT_sim}-(a) shows that $\alpha_2^{*}(\varphi ; d)$ 
from HS simulations seem to saturate to $\alpha_2^{\mathrm{sat.}}(d)$ as $\varepsilon$ decreases, i.e.\ as the (avoided) critical point $\varphi_c$ is approached. The trend becomes more robust as $d$ increases. In addition, we find that the total magnitude of $\alpha_2^*(\varphi ; d)$ decreases with increasing $d$. These two observations suggest that collective effects diminish and the physics becomes more single particle-like with increasing $d$, consistent with expectations from DMFT. Figure~\ref{fig:NGP_peak_scaling_MCT_sim}-(a) also shows the growth of the NGP peak with the appropriate exponent [$\alpha_2^*(\varphi ; d) \propto \varepsilon^{-1}$] in a finite $\varepsilon$ range. This again confirms the finite-$d$ relevance of the $d\rightarrow\infty$ description. We find that MCT also predicts a clear saturation of $\alpha_2^*(\varphi ; d)$ to some finite value in the ergodic phase (see Fig.~\ref{fig:NGP_peak_scaling_MCT_sim}-(b)). However, the MCT-predicted saturation height \emph{shrinks} with increasing $d$. We also note that the linear divergence of the saturation value $\alpha_2^{\mathrm{sat.}}(d)$ with $d$, while evident for the simulated MK model, is less clear for the simulated HS system (see Fig.~\ref{fig:NGP_sat_HS_MK}). This can partly be attributed to (i) the uncertainty in determining the location of the (pseudo) critical point $\varphi_c(d)$ which stems from the computationally demanding equilibration procedures required for the latter system, as well as (ii) collective contributions to the NGP in the HS system. Simulations at even lower $\varepsilon$ (and also higher $d$) would be necessary to obtain a more reliable estimate of the plateau height.

\begin{figure}[ht]
    \centering
    \includegraphics[width=0.75\columnwidth]{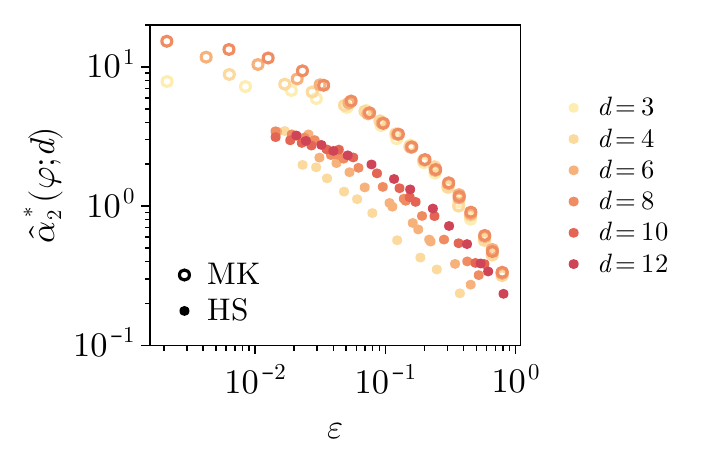}
    \caption{Peak of the (dimensionally scaled) NGP from simulations of HS (closed circles) and  MK model (open circles) simulations. Data for the MK model are reproduced from Ref.~\citenum{charbonneau2024dynamics}.}
\label{fig:HS_sim_NGP_sat}
\end{figure}

A key observation of the present study is that both numerical simulations and MCT calculations suggest that structural features reduce single-particle fluctuations. Specifically, we find that the dimensionally rescaled NGP peak $\hat{\alpha}_2^*(\varphi ; d)$ for the MK model is consistently larger than for the HS system, as shown in Fig.~\ref{fig:HS_sim_NGP_sat}-(a). More precisely, the NGP peak height for the MK model -- across all dimensions studied -- is larger by a factor of 5--8 than that for HS. The MCT calculations in Fig.~\ref{fig:qualitative_comparison_MCT_DMFT} (e)-(f) also show a difference of a factor 2--3 between the two models. We hypothesize that the more prominent local structure of the HS system might be responsible for this difference. Given that MCT captures the quantitative difference between the two systems based solely on pair-structure information, we conjecture that pair-structure likely plays a significant role.  It is nonetheless uncertain whether the reduction of the strength of the critical fluctuations is entirely due to pair structure or if higher-order structural correlations\cite{zhang2020revealing, luo2022many, pihlajamaa2023emergent} also contribute significantly to this effect. In any event, these findings suggest that while structural features \emph{enhance} collective fluctuations (as measured from four-point functions, see Refs.~\citenum{mari2011dynamical, nishikawa2022collective}), they simultaneously \emph{reduce} single-particle pseudo-critical fluctuations in the supercooled state.

\section{Conclusion}
\label{sec:conclusion}
Motivated by testing theories of the glass transition on recently identified pseudo-critical single-particle fluctuations, we provide a theoretical and numerical analysis of both MCT and the more recent DMFT for simple glass formers. We first derive general scaling results which any theory of liquid dynamics should satisfy in the limit $d\rightarrow\infty$, and find that the standard MCT is incompatible with these scalings,  unlike the DMFT (see Table \ref{tab:scaling_summary}). 
The inability of MCT  to predict appropriate dimensional scalings stems from the inherent dimensionally dependent UV cutoff of the theory. MCT is nevertheless found to correctly capture the scope of the mean-field dynamical universality class of hard spheres. By enforcing a Gaussian ansatz for the non-ergodicity parameter on MCT, we also show that the correct dimensional scalings are recovered (up to multiplicative pre-factors).

Additionally, analysis of the dimensional dependence of the critical exponent $\gamma$ controlling the diverging timescale associated with single particle critical fluctuations (as well as the vanishing of the diffusivity) in hard sphere systems shows a monotonic convergence to the appropriate large $d$ asymptotic value. Our results further reveal that MCT manages to capture the expected dimensional trends, such as the saturation of the peak of the non-Gaussian parameter as $|\varphi-\varphi_c| \rightarrow0$ in moderate $d$. Numerical solutions of the MCT equations  confirm the dimensional scalings we have derived
%from Sec.~\ref{sec:general_NGP}, 
as solutions for both the (dimensionally rescaled) mean squared displacement and non-Gaussian parameter appear to converge to a master curve as $d$ increases.  

Comparisons of MCT results at finite times with state-of-the-art simulations of hard spheres for $4\leq d\leq12$ provide additional insight. While the predictions for the mean squared displacement are qualitatively similar, the predictions for the non-Gaussian parameter differ significantly. Simulation results for the non-Gaussian parameter are always positive, whereas MCT predicts it to turn negative as $d$ increases. We believe that this discrepancy is rooted in the unphysical predictions of the mean quartic displacement and the magnitude of the theory's `critical amplitudes'.\cite{gotze2009complex} Future studies should focus on verifying this hypothesis. In moderately low $d$ MCT nevertheless still predicts correctly that the non-Gaussian parameter first grows and peaks before vanishing in the diffusive regime.

Lastly, we have considered the effect of local structure on single particle critical dynamical fluctuations by comparing results from hard spheres and the Mari--Kurchan model. The notion that fluid structure is central to explaining the origins of the dramatic dynamical slowing down in finite $d$ liquid glass formers is not new.\cite{royall2015role, singh2023intermediate} 
Our findings reveal that a more pronounced fluid structure (like in hard spheres) diminishes the magnitude of single-particle fluctuations, whereas other studies\cite{mari2011dynamical, adhikari2021} have reported that it should increase collective ones. While we cannot exclude that higher-order structural features also matter at low $d$, our mode-coupling analysis leads us to hypothesize that pair structure plays a more dominant role in explaining this confounding observation. Overall, however, identifying the effect of local structure on simple fluctuations remains a largely open problem.

The conclusions drawn from this work highlight certain limitations of MCT concerning simple fluctuations, especially in the limit of growing spatial dimensions, but also in finite $d$ systems. By contrast, DMFT -- and its physically expected finite-$d$ corrections -- is compatible with observations from numerical simulations. These findings, however, should not detract from the numerous successes and significant insights into glass formation that MCT provides. In particular, we stress that MCT, unlike the existing (single-particle) DMFT, predicts the slowing down of equilibrium relaxation of \emph{collective density fluctuations}.\cite{gotze2009complex} MCT further predicts diverging fluctuations of a collective origin -- distinct from single-particle ones -- as dynamical arrest is approached.\cite{biroli2006inhomogeneous} These fluctuations which correspond to non-linear responses to infinitesimal, static density modulations, have been used to measure growing correlation lengths in finite-dimensional supercooled liquids.\cite{berthier2007spontaneousII, kim2013dynamic, albert2016fifth} Given this context, a dedicated effort to resolving the unphysical features of MCT identified in this work is therefore amply worth the effort. The result could well be an integral part of the definitive theory of glasses.

\paragraph*{Author Contributions:} CCLL: Conceptualization, Methodology, Investigation \& Formal Analysis, Visualization, Writing (Original Draft); PC: Conceptualization, Methodology, Investigation \& Formal Analysis, Writing (Original Draft); YH: Methodology, Investigation \& Formal Analysis, Writing (Review \& Editing); LMCJ: Conceptualization, Writing (Review \& Editing); PM: Conceptualization, Writing (Review \& Editing); IP: Conceptualization, Software, Writing (Review \& Editing); GS: Conceptualization, Methodology, Writing (Review \& Editing).

\begin{acknowledgement}
PC acknowledges support from the Simons Foundation (Grant No.~454937), and thanks the Chimera group of Sapienza for hosting his sabbatical year during which this work was initiated. CCLL, IP, and LMCJ acknowledge financial support from the Dutch Research Council (NWO) through a Vidi grant. GS gratefully acknowledges the support of NSF Grant No.~CHE 2154241.
\end{acknowledgement}

\appendix
\section{Numerical Details for MCT Calculations \& Additional Results}

\subsection{Phase Diagram}
To compute the phase diagram of HS, we solve the following self-consistent equation for the non-ergodicity parameter $\phi^{\infty}(q)$:
    \begin{equation}
        \frac{\phi^{\infty}(q)}{1-\phi^{\infty}(q)} = \frac{\rho_0D_0 S(q)}{q^2}\int\frac{\mathrm{d}\boldsymbol{k}}{(2\pi)^d}\ V(\boldsymbol{q},\boldsymbol{k})^2 \phi^{\infty}(k)\phi^{\infty}(|\boldsymbol{q}-\boldsymbol{k}|)
    \end{equation}
where 
    \begin{equation}
    \begin{split}
        V(\boldsymbol{q},\boldsymbol{k})^2 = (\hat{\boldsymbol{q}}\cdot\boldsymbol{k})& c(q) \left[ (\hat{\boldsymbol{q}}\cdot\boldsymbol{k})c(k) + \left(\hat{\boldsymbol{q}}\cdot(\boldsymbol{q}-\boldsymbol{k})\right) c(|\boldsymbol{q}-\boldsymbol{k}|)\right]\\
        &\times S(k)S(|\boldsymbol{q}-\boldsymbol{k}|),
    \end{split}
    \label{eq:MCT_eos}
    \end{equation}
which is valid for HS fluids. The microscopic HS $S(k)$ is obtained by solving the PY equations using a recent open-source implementation that provides solutions to the Ornstein-Zernike equation,\cite{pihlajamaaOZ} in which we made use of the recursive iteration in Fourier space with the Ng acceleration method.\cite{ng1974hypernetted} A similar equation of state can be derived for the MK model: 
    \begin{equation}
        \frac{\phi^{\infty}_s(q)}{1-\phi_s^{\infty}(q)} = \frac{D_0\rho_0}{q^2} \int \frac{\mathrm{d}\boldsymbol{k}}{(2\pi)^d} (\boldsymbol{q}\cdot\boldsymbol{k})^2 c(k)^2\phi_s^{\infty}(k)\phi_s^{\infty}(|\boldsymbol{q}-\boldsymbol{k}|).
    \end{equation}

We perform a bisection search in $\rho_0$ (or equivalently the dimensionally rescaled packing fraction $\tilde{\varphi}$) to find non-trivial solutions $\phi^{\infty}(q)$ to Eq.~\eqref{eq:MCT_eos}. Following prior studies on MCT, we express the integral in bispherical coordinates and use a uniform grid of $N_k=200$ on a uniform wave-vector grid $k\in(0, k_{\mathrm{max}})$ where $k_{\mathrm{max}}$ is chosen such that sufficient oscillations in the direct correlation function are taken into account, leading to a converged result. 

\subsection{Dynamics}

We solve the dynamical MCT equations using a logarithmically coarse-grained time grid,\cite{fuchs1991comments} which has recently been implemented in an open-source solver for integro-differential equations of mode-coupling type.\cite{pihlajamaa2023modecouplingtheory} Akin to the static calculations, the wave-vector integrals were performed (in bispherical coordinates) over a grid consisting of $N_k=200$ points in the interval $k\in(0,60)$. This choice was checked to be sufficient to ensure numerical convergence in $d\leq20$. Note that for these results, we have re-computed the critical points with these settings to greater precision, in order to allow for solutions at small relative distances to them.

\subsection{Transient Negative Behavior of the MQD}

\begin{figure}[h]
    \centering
\includegraphics[width=\columnwidth]{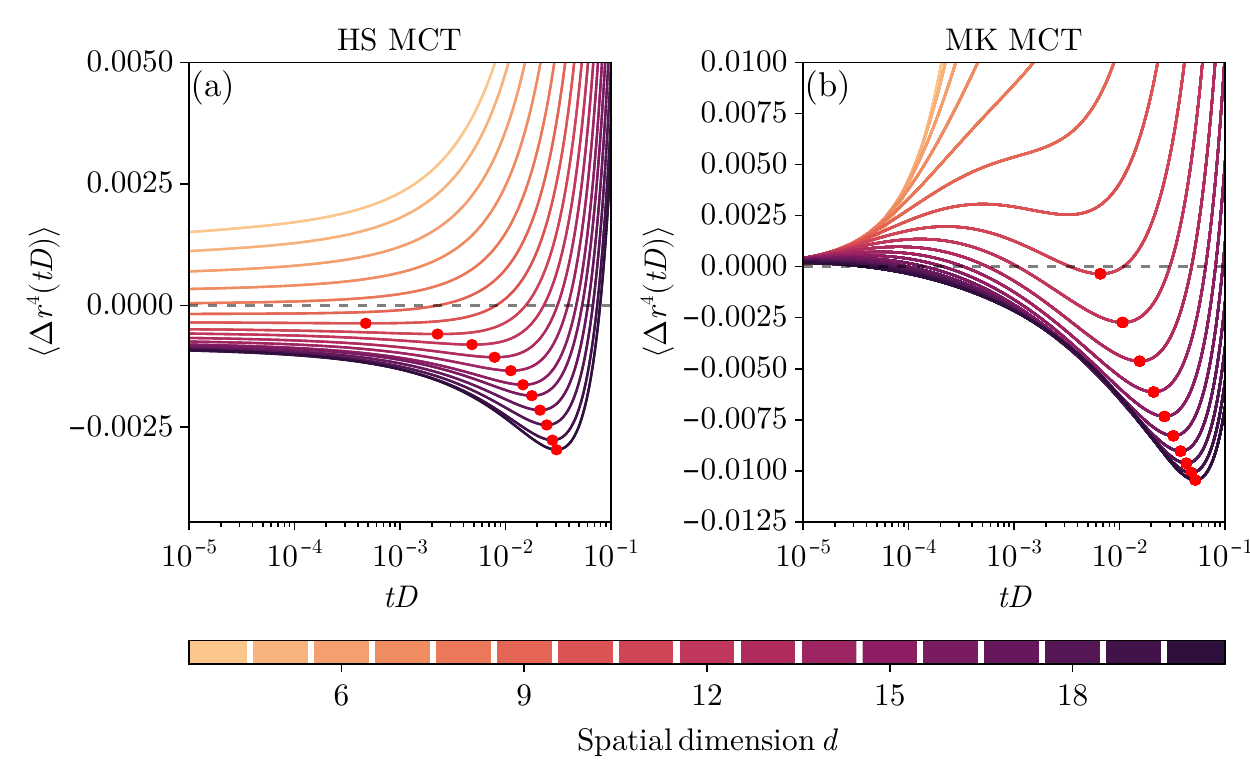}
    \caption{Illustration of the negative dips of the mean quartic displacements that MCT predicts. Red dots indicate the negative minimum of the curves.}
    \label{fig:negative_dip_MQD}
\end{figure}

As discussed in the main text, our results show a complete breakdown of MCT which predicts transient negative dips of the MQD for all dimensions $d\geq8$. We do not believe these to be numerical artifacts, but instead genuine properties of the solutions. We have checked that increasing numerical the numerical accuracy of the integration methods does not remedy this problematic prediction. We show in Fig.~\ref{fig:negative_dip_MQD} examples of this behavior near the dynamical transition for both HS and the MK model. The red dots indicate the minimum of the curves, we see that the dip size increases with dimension, it is unclear whether or not it will diverge to negative infinity in the limit $d\rightarrow\infty$. We emphasize once more that similar negative dips have been observed in mode-coupling solutions for the self part of the Van Hove function at long times ($G_s(r,t\rightarrow\infty)$), which was inherently linked to the non-Gaussian form of the non-ergodicity parameter predicted by the theory.\cite{ikeda2010, ikeda2011ikedareply} While unfortunate, these results thus remain consistent with earlier studies.

\subsection{Scaling Function $G(t)$ and Associated Dynamical Exponents}
\label{app:dynamical_exponents}

The scaling function $G(t)$ introduced in Eq.~\eqref{eq:NGP_asymptotic} is a central prediction of MCT. While we mentioned that it describes the growth of the NGP in moderately low dimensions, it first and foremost describes the behavior of the MSD around the caging regime: $\langle \Delta r^2(t) \rangle = 2d R - h_{\mathrm{MSD}}G(t) - h_{\mathrm{MSD}}[H(t) + K_{\mathrm{MSD}}G(t)^2 + \varepsilon \hat{K}_{\mathrm{MSD}}]$, where critical amplitudes analogous to those defined for the NGP have been introduced.\cite{fuchs1998asymptotic} The functions $G(t)$ and $H(t)$ are however the same, a property known as the `factorization theorem' in the mode-coupling literature.\cite{gotze1999recent} It is possible to show that the leading order scaling function $G(t)$ behaves as $G(\hat{t}\ll1) \approx \hat{t}^{-a}$, $G(\hat{t}\approx1) \approx 1$ and  $G(\hat{t}\gg1) \approx -\hat{t}^{b}$ with $\hat{t} = t / t_{\varepsilon}$, where  $a,\ b$ are two dynamical critical exponents. \cite{fuchs1998asymptotic, gotze2009complex} The timescale $t_{\varepsilon}$ is a free parameter (around which the expansion for the asymptotic dynamics in the caging regime is originally performed) and satisfies $t_0 < t_{\varepsilon} < t_d$, where $t_0$ marks the end of the short-time behavior and $t_d$ corresponds to the crossover to the long-time limit and thus diverges at criticality, when $\varepsilon\rightarrow0$.\cite{franosch1994theory} Furthermore, the two dynamical exponents in fact define the critical exponent $\gamma$ discussed in the main text
    \begin{equation}
        \gamma = \frac{1}{2a} + \frac{1}{2b}.
    \label{eq:gamma_def}
    \end{equation}
One can also show that they are related by the following 
    \begin{equation}
        \frac{\Gamma(1-a)^2}{\Gamma(1-2a)} = \frac{\Gamma(1+b)^2}{\Gamma(1+2b)}
    \label{eq:a_b_relation}
    \end{equation}
where $\Gamma(x)$ denotes the Euler-$\Gamma$ function. 

We next study the dimensional dependence of the dynamical exponents $a$ and $b$. Once $\gamma$ has been obtained by the methods described in the main text, we then use Eqs.~\eqref{eq:gamma_def}-\eqref{eq:a_b_relation} above to numerically solve for $a,b$. Akin to the results for the critical exponent $\gamma$, we find that results from our mode-coupling calculations and from the computer simulations converge to their respective limits, with the exception of the measurements from the diverging timescale associated with critical fluctuations in simulations for which convergence is obstructed, particularly in the HS system. Reasons for this disagreement have been proposed in the main text and extend to the present analysis as well. 

\begin{figure}[h]
    \centering
\includegraphics[width=0.75\columnwidth]{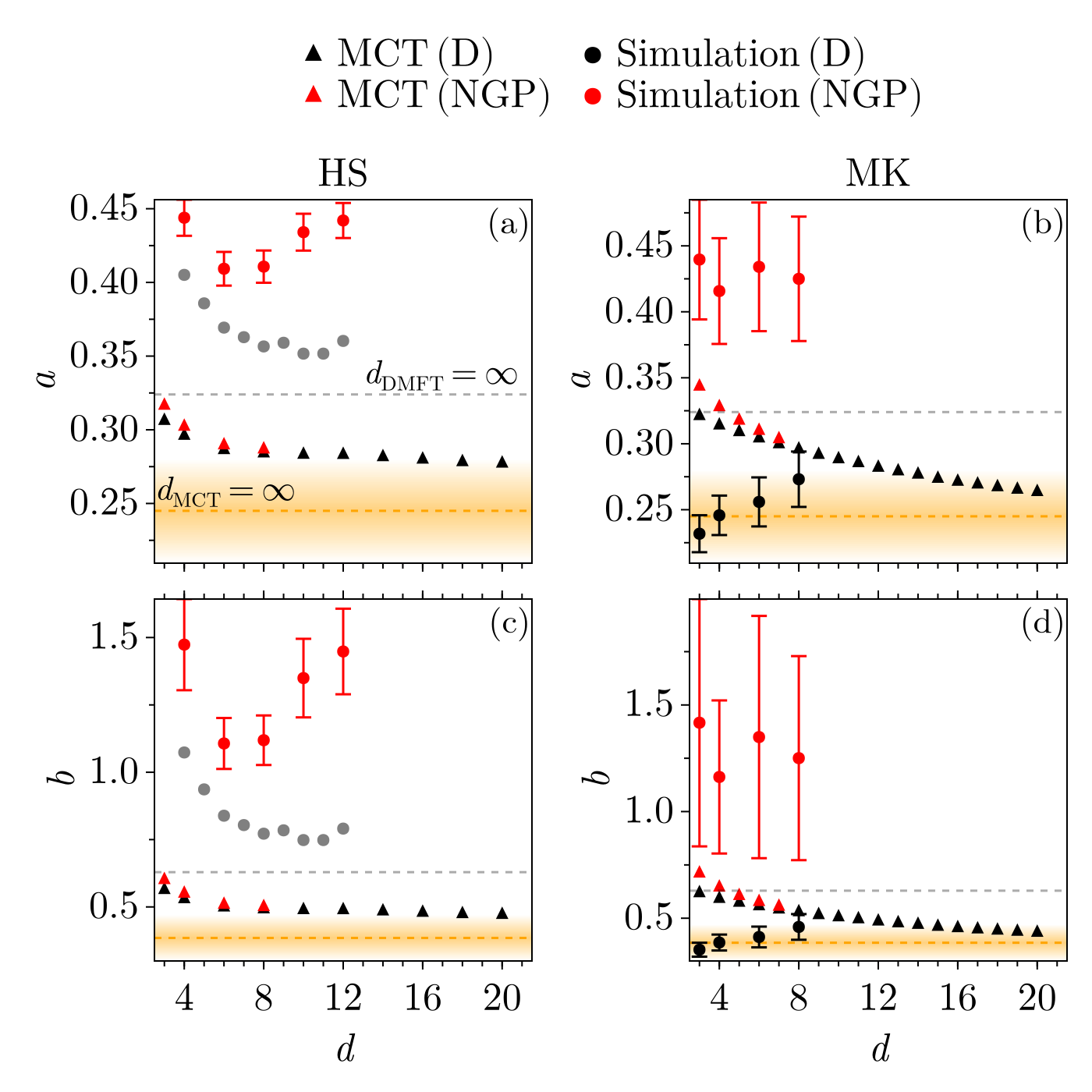}
    \caption{Dynamical exponents $a$, $b$ governing the early and late $\beta$-regime as a function of dimension $d$ for HS and the MK model.}
    \label{fig:dynamical_exponents_a_b_HS_MK}
\end{figure}

\clearpage

\bibliography{fluctuations}% Produces the bibliography via BibTeX.

\end{document}

\title{Supplementary Material: Simple fluctuations in simple glass formers}

\maketitle

%We provide below details of the calculations omitted in the main text. 

\section{General Derivation of the MSD \& the MQD}
As mentioned in the main text, we can derive the equation of motion for the NGP by computing the second and fourth moment of the displacement distributions. The self-intermediate scattering function (ISF) is related to the self Van Hove function $G_s(\boldsymbol{r},t)$ (see main text for definition) by a spatial Fourier transform: 
    \begin{equation}
    \begin{split}
        \phi_s(\boldsymbol{q},t) =&\ \int_{\mathbb{R}^d} \mathrm{d}\boldsymbol{r}\  e^{-i\boldsymbol{q}\cdot\boldsymbol{r}}G_s(\boldsymbol{r},t) \\
        =&\ 1 - \frac{1}{2!}\int_{\mathbb{R}^d} \mathrm{d}\boldsymbol{r}\  q^2r^2\cos^2(\theta)G_s(\boldsymbol{r},t) + \frac{1}{4!}\int_{\mathbb{R}^d} \mathrm{d}\boldsymbol{r}\ q^4r^4\cos^4(\theta)G_s(\boldsymbol{r},t) + \mathcal{O}(q^6)
    \end{split}
    \label{eq:expansion_ISF}
    \end{equation}
where there are no odd-terms in $r \equiv |\boldsymbol{r}|$ since we have a non-chiral system. Here, $\theta$ denotes the angle between the vectors $\boldsymbol{r}$ and $\boldsymbol{q}$. By isotropy we can write $G_s(\boldsymbol{r},t) = G_s(r,t)$, and thus $\phi_s(q,t)$ where $r \equiv |\boldsymbol{r}|$ and $q \equiv |\boldsymbol{q}|$. By definition, the MSD is equal to the second moment of $G(r,t)$. Hence we have 
    \begin{equation}
        \int_{\mathbb{R}^d} \mathrm{d}\boldsymbol{r}\  r^2G_s(r,t) =\ \Omega_d\int_0^{\infty} \mathrm{d}r\ r^{d-1} r^2 G_s(r,t) \equiv \langle \Delta r^2(t) \rangle.
    \end{equation}
We can then write the second term of Eq.~\eqref{eq:expansion_ISF}
    \begin{equation}
    \begin{split}
        \int_{\mathbb{R}^d} \mathrm{d}\boldsymbol{r}\  r^2\cos^2(\theta)G_s(\boldsymbol{r},t)
        =&\ \frac{\Omega_d}{d}\int_0^{\infty}\mathrm{d}r\  r^{d-1} r^2 G_s(r,t) \\
        =&\ \frac{\langle \Delta r^2(t)\rangle}{d}.
    \end{split}
    \end{equation}
Similarly we have using the definition of the fourth moment: 
    \begin{equation}
        \langle \Delta r^4(t)\rangle \equiv \int_{\mathbb{R}^d} d\boldsymbol{r}\ r^4G_s(r,t),
    \end{equation}
and thus that the third term of Eq.~\eqref{eq:expansion_ISF} reads
    \begin{equation}
    \begin{split}
        \int_{\mathbb{R}^d} \mathrm{d}\boldsymbol{r}\ r^4\cos^4(\theta)G_s(\boldsymbol{r},t)
        =&\ \frac{3\Omega_d}{d(d+2)}\int_0^{\infty}\mathrm{d}r\  r^{d-1}\ r^{4}G_s(r,t)\\
        =&\ \frac{3\langle \Delta r^4(t)\rangle}{d(d+2)}.
    \end{split}
    \end{equation}
We can then write down the low-$q$ expansion of the tagged ISF as
    \begin{equation}
        \phi_s(q,t) \approx 1 - \frac{q^2}{2d}\langle \Delta r^2(t)\rangle + \frac{3q^4}{4!d(d+2)}\langle \Delta r^4(t)\rangle + \mathcal{O}(q^6).
    \end{equation}
We then write down $q^2m_s(q,t) = m_0(t) + q^2 m_2(t)/2$ and expand the general equation for $\phi_s(q,t)$ (i.e. Eq.~(6) in the main text). We get to $\mathcal{O}(q^4)$:
    \begin{equation}
    \begin{split}
        0 =&\frac{\mathrm{d}}{\mathrm{d}t}\left[- \frac{q^2}{2d}\langle \Delta r^2(t)\rangle + \frac{3q^4}{4!d(d+2)}\langle \Delta r^4(t)\rangle \right] + q^2D_0\left(1 - \frac{q^2}{2d}\langle \Delta r^2(t)\rangle\right)\\
    &+ D_0\int_0^t\mathrm{d}\tau\left(m_0(t-\tau) + \frac{q^2}{2} \ m_2(t-\tau) \right)\frac{\mathrm{d}}{\mathrm{d}\tau}\left[- \frac{q^2}{2d}\langle \Delta r^2(\tau)\rangle + \frac{3q^4}{4!d(d+2)}\langle \Delta r^4(\tau)\rangle \right].
    \end{split}
    \end{equation}
Matching terms of $\mathcal{O}(q^2)$ and $\mathcal{O}(q^4)$ respectively gives the equations of motion for the MSD [Eq.~(7)] and the MQD [Eq.~(8)] presented in the main text. 

\section{Low-$q$ Behavior of Tagged Memory Kernel from MCT}
In $d$ dimensions, the MCT describes single particle dynamics via the self-ISF
\begin{equation}
        \gamma_q \dot{\phi}_s(q,t) + \phi_s(q,t) + \int_0^t \mathrm{d}\tau m_s(q, t-\tau)\dot{\phi}_s(q,\tau) = 0
    \label{eq:tagged_MCT}
    \end{equation}
where $\gamma_q \equiv (q^2D_0)^{-1}$ sets the natural timescale ($D_0$ denotes the bare diffusion constant) and with a memory kernel approximated by
\begin{equation}
\begin{split}
m_s(q,t) &= \frac{n}{q^4}\int_{\mathbb{R}^d} \frac{\mathrm{d}\boldsymbol{k}}{(2\pi)^d}(\boldsymbol{q}\cdot\boldsymbol{k})^2c(k)^2\phi(k,t)\phi_s(|\boldsymbol{q}-\boldsymbol{k}|,t), 
\end{split}
\end{equation}
where $\phi(k,t),\ \phi_s(k,t)$ denote the full and tagged intermediate scattering functions and $c(k)$ the direct correlation function. As stated above, we want to write down the low-momentum behavior of $m_s(q,t)$ to second order around $q=0$, which gives
    \begin{equation}
        m_s(q,t) = m_s(q,t)|_{q=0} + \frac{1}{2}q^2\pdv{^2m_s(q,t)}{q^2}\bigg\vert_{q=0} + \mathcal{O}(q^4).
    \end{equation}
We note, again, that there are no even terms since we have a non-chiral system. In spherical coordinates, the coefficients are given by 
    \begin{equation}
    \begin{split}
        m_s(q,t)|_{q=0} &= \frac{n}{q^2(2\pi)^d}\int_0^{\infty}\mathrm{d}k \int_0^{\pi}\mathrm{d}\theta_1...\mathrm{d}\theta_{d-2}\int_0^{2\pi}\mathrm{d}\varphi \sin^{d-2}(\theta_1)\sin^{d-3}(\theta_2)...\sin(\theta_{d-2})k^{d+1}\cos^2(\theta_1)c(k)^2\phi_{k}(t)\phi_{k}^{(s)}(t)\\
        % &= \frac{n}{q^2(2\pi)^d}\frac{2\pi^{(d-1)/2}}{\Gamma((d-1)/2)}\int_0^{\infty}dk k^{d+1}c(k)^2\phi_{k}(t)\phi_{k}^{(s)}(t) \int_0^{\pi}d\theta_1 \sin^{d-2}(\theta_1)\cos^2(\theta_1) \\
        &= \frac{n}{(2\pi)^d}\frac{\Omega_{d}}{d}\frac{1}{q^2}\int_0^{\infty} \mathrm{d}k k^{d+1}S(k)c(k)^2\phi_k(t)\phi_s(k,t)
    \end{split}
    \end{equation}
and
\begin{equation}    
\begin{split}
        \pdv{^2m_s(q,t)}{q^2}\bigg\vert_{q=0}&=\lim_{q\rightarrow0}\frac{n}{(2\pi)^d}\int_0^{\infty}\mathrm{d}k\int_0^{\pi}\mathrm{d}\theta_1...\mathrm{d}\theta_{d-2}\int_0^{2\pi}\mathrm{d}\varphi J(k,\{\theta_i\})k^2\cos^2(\theta_1)S(k)c(k)^2\phi_k(t)\partial_q^2[q^{-2}\phi_{|\boldsymbol{q}-\boldsymbol{k}|}^{(s)}(t)].
    \end{split}
    \end{equation}
where $J(k,\{\theta_i\})$ denotes the spherical Jacobian. We then consider 
    \begin{equation}
    \begin{split}
        \partial_q^2[q^{-2}\phi_{|\boldsymbol{q}-\boldsymbol{k}|}^{(s)}(t)]|_{q\rightarrow0} &= \left[6q^{-4}\phi_s(p,t) - 4q^{-3}\pdv{p}{q}\pdv{\phi_s(p,t)}{p} + q^{-2}\pdv{^2p}{q^2}\pdv{^2\phi_s(p,t)}{p^2} + q^{-2}\left(\pdv{p}{q}\right)^2\pdv{^2\phi_s(p,t)}{p^2} \right]\Bigg|_{q\rightarrow0} \\
        % &=q^{-2}\left(\pdv{^2p}{q^2} \pdv{\phi_s(p,t)}{p} + \left(\pdv{p}{q} \right)^2 \pdv{^2\phi_s(p,t)}{p^2}\right)\Bigg|_{q\rightarrow0} + \mathcal{O}(q^{-3}) \\
        &= q^{-2}\left(\frac{\sin^2(\theta_1)}{k}\pdv{\phi_s(k,t)}{k} + \cos^2(\theta_1)\pdv{^2\phi_s(k,t)}{k^2} \right)+\mathcal{O}(q^{-3})
    \end{split}
    \end{equation}
where we have used $ |\boldsymbol{q}-\boldsymbol{k}| \equiv p = \sqrt{q^2+k^2-2kq\cos(\theta)} \approx k - \cos(\theta)q + \frac{1}{2}k^{-1}\sin^2(\theta_1)q^2 + \mathcal{O}(q^3)$. Integrating over the angular variables, we can show that
    \begin{equation}
    \begin{split}
        \pdv{^2m_s(q,t)}{q^2}\bigg\vert_{q=0} &= \frac{n}{(2\pi)^d q^2}\frac{2\pi^{(d-1)/2}}{\Gamma((d-1)/2)}\int_0^{\infty}\mathrm{d}k \int_0^{\pi}\mathrm{d}\theta_1 k^{d+1} \sin^{d-2}(\theta_1)\cos^2(\theta_1)S(k)c(k)^2\phi_k(t)\\
        &\hspace{5cm}\times \ \left(\frac{\sin^2(\theta_1)}{k}\pdv{\phi_s(k,t)}{k} + \cos^2(\theta_1)\pdv{^2\phi_s(k,t)}{k^2} \right) \\
        % &= \frac{n}{(2\pi)^d}\frac{2\pi^{(d-1)/2}}{\Gamma((d-1)/2)}\int_0^{\infty}dk k^{d+1}S(k)c(k)^2\phi_k(t) \left(\frac{1}{k}\frac{\sqrt{\pi}\Gamma((d+1)/2)}{2\Gamma((4+d)/2)}\pdv{\phi_s(k,t)}{k} + \frac{3\sqrt{\pi}\Gamma((d-1)/2)}{4\Gamma((d+4)/2)}\pdv{^2\phi_s(k,t)}{k^2} \right) \\
        % &= \frac{n}{2(2\pi)^d}\frac{2\pi^{d/2}}{\Gamma((d-1)/2)} \frac{1}{\Gamma((d+4)/2)} \int_0^{\infty}dk k^{d+1}S(k)c(k)^2 \phi_k(t)\left(\frac{\Gamma((d+1)/2)}{k}\pdv{\phi_s(k,t)}{k} + \frac{3\Gamma((d-1)/2)}{2}\pdv{^2\phi_s(k,t)}{k^2} \right)
        &=\frac{1}{q^2}\frac{3n \Omega_d}{(2\pi)^d d(d+2)} \int_0^{\infty}\mathrm{d}k k^{d+1} S(k)c(k)^2\phi_k(t)\left(\frac{d-1}{3k}\pdv{\phi_s(k,t)}{k} + \pdv{^2\phi_s(k,t)}{k^2} \right).
    \end{split} 
    \end{equation}
% where we have used 
%     \begin{equation}
%         \int_0^{\pi}d\theta\sin^{d}(\theta)\cos^2(\theta) = \frac{\sqrt{\pi}\Gamma((d+1)/2)}{2\Gamma((4+d)/2)}
%     \end{equation}
% and 
%     \begin{equation}
%         \int_0^{\pi}d\theta\sin^{d-2}(\theta)\cos^4(\theta) = \frac{3\sqrt{\pi}\Gamma((d-1)/2)}{4\Gamma((d+4)/2)}.
%     \end{equation}
% We can simplify things further by multiplying through by $\Gamma(d/2) / \Gamma(d/2)$ and distributing the factor $\Gamma((d-1)/2)^{-1}$ inside of the square brackets. Then using 
%     \begin{equation}
%         \frac{\Gamma((d+1)/2)}{\Gamma((d-1)/2)} = \frac{d-1}{2}
%     \end{equation}
% and 
%     \begin{equation}
%         \frac{\Gamma(d/2)}{\Gamma((d+4)/2)} = \frac{4}{2d+d^2},
%     \end{equation}
% we eventually obtain
%     \begin{equation}
%         m_2^{(s)}(t) = \frac{1}{q^2}\frac{3n \Omega_d}{(2\pi)^d d(d+2)} \int_0^{\infty}dk k^{d+1} S(k)c(k)^2\phi_k(t)\left(\frac{d-1}{3k}\pdv{\phi_s(k,t)}{k} + \pdv{^2\phi_s(k,t)}{k^2} \right).
%     \end{equation}
As a sanity check, substituting $d=3$, we recover the result of Flenner \& Szamel, see Eq.~(27) of \cite{flenner2005relaxation}:
     \begin{equation}
        \pdv{^2m_s(q,t)}{q^2}\bigg\vert_{q=0} =  \frac{1}{q^2}\frac{n}{10\pi^2}\int_0^{\infty}\mathrm{d}k k^4S(k)c(k)^2\phi_k(t)\left(\frac{2}{3k}\pdv{\phi_s(k,t)}{k} + \pdv{^2\phi_s(k,t)}{k^2}  \right).
    \end{equation}

\section{Dimensional Scalings of Observables within MCT}
We proceed with the details of the calculations for the various dimensional scalings within MCT whose results have been presented and discussed in the main text.

\subsection{Microscopic MCT}
We demonstrated above that the equations of motion for the MSD and the MQD are given by 
    \begin{equation}
        \frac{\mathrm{d}\langle\Delta r^2(t)\rangle}{\mathrm{d}t} + D_0\int_0^t\mathrm{d}\tau m_0(t-\tau)\frac{\mathrm{d}\langle\Delta r^2(\tau)\rangle}{\mathrm{d}\tau} = 2dD_0
    \label{eq:MSD}
    \end{equation}
and 
    \begin{equation}
        \frac{\mathrm{d}\langle \Delta r^4(t)\rangle}{\mathrm{d}t}- 4(d+2)D_0\langle \Delta r^2(t) \rangle + D_0 \int_0^t \mathrm{d}\tau m_0(t-\tau)\frac{\mathrm{d}\langle \Delta r^4(\tau)\rangle}{\mathrm{d}\tau}-2(d+2)D_0\int_0^t \mathrm{d}\tau m_2(t-\tau)\frac{\mathrm{d}\langle \Delta r^2(\tau)\rangle}{\mathrm{d}\tau} = 0
    \label{eq:MQD}
    \end{equation}
where 
    \begin{equation}
        m_0(t) = \frac{n}{(2\pi)^d}\frac{\Omega_{d}}{d}\int_0^{\infty} \mathrm{d}k k^{d+1}S(k)c(k)^2\phi_k(t)\phi_s(k,t)
    \end{equation}
and 
    \begin{equation}
        m_2(t) = \frac{3n \Omega_d}{(2\pi)^d d(d+2)} \int_0^{\infty}\mathrm{d}k k^{d+1} S(k)c(k)^2\phi_k(t)\left(\frac{d-1}{3k}\pdv{\phi_s(k,t)}{k} + \pdv{^2\phi_s(k,t)}{k^2} \right).
    \end{equation}
To study in the limit of large $d$,  follow the procedure outlined in Schmid \& Schilling \cite{schmid2010}. We must first re-express all quantities over the scale $k = \tilde{k}d$. We can then  make use of 
    \begin{equation}
        S(\tilde{k}d) \approx 1 
    \end{equation}
and 
    \begin{equation}
        c(\tilde{k}d)^2 \approx 2 (2\pi)^d d^{-d} \left[\tilde{k}^d \pi d \sqrt{(2\tilde{k})^2-1} \right]^{-1} \times \Theta(\frac{1}{2}-\tilde{k}).
    \end{equation}
Lastly, we must use the dimensional scaling of the critical packing fraction, which for MCT will scale as $\varphi \equiv nV_d = n\Omega_d / (2^d d) = \tilde{\varphi} \times d^2 2^{-d}$ along with the non-ergodicity parameter $f_s(\tilde{k}d) = f(\tilde{k}d) \equiv \tilde{f}(\tilde{k})\approx \Theta(\tilde{k}_0 - \tilde{k})$ with $\tilde{k}_0 = x\sqrt{d}$ for some numerical pre-factor $x$. As mentioned in the main text, we perform our analysis in the non-ergodic phase, where analytic results presented above are known to hold. 

\paragraph{Scaling of $m_0$}: in the non-ergodic phase we get 
    \begin{equation}
    \begin{split}
        m_0 =&\  \frac{n}{(2\pi)^d}\frac{\Omega_{d}}{d}\int_0^{\infty} \mathrm{d}(\tilde{k}d) (\tilde{k}d)^{d+1}c(\tilde{k}d)^2f(\tilde{k}d)^2 \\
        % =&\  \frac{n}{(2\pi)^d}\frac{\Omega_{d}}{d}\int_0^{\infty} d(\tilde{k}d) (\tilde{k}d)^{d+1}\left[ 2 (2\pi)^d d^{-d} \left[\tilde{k}^d \pi d \sqrt{(2\tilde{k})^2-1} \right]^{-1} \times \Theta(\frac{1}{2}-\tilde{k})\right]\Theta(\tilde{k}_0 - \tilde{k})^2 \\
        % =&\ \frac{2n\Omega_d}{\pi}\int_{1/2}^{\infty} d\tilde{k} \frac{\tilde{k}}{\sqrt{(2\tilde{k})^2-1}}\Theta(x\sqrt{d}-\tilde{k})\\
        =&\ \frac{2 \tilde{\varphi}d^3}{\pi}\int_{1/2}^{x\sqrt{d}} \mathrm{d}\tilde{k} \frac{\tilde{k}}{\sqrt{(2\tilde{k})^2-1}} \\
        =&\ d^{7/2} \times \frac{\tilde{\varphi}}{2\pi}\sqrt{4x^2 - \frac{1}{d}}.
    \end{split}
    \end{equation}
Thus we find that $m_0 \sim d^{\gamma}$ with $\gamma = 7/2$.
\paragraph{Scaling of $m_2$}: in the non-ergodic phase we get 
    \begin{equation}
    \begin{split}
        m_2 =&\ \frac{3n \Omega_d}{(2\pi)^d d(d+2)} \int_0^{\infty}\mathrm{d}(\tilde{k}d) (\tilde{k}d)^{d+1} c(\tilde{k}d)^2f(\tilde{k}d)\left(\frac{d-1}{3(\tilde{k}d)}\frac{\mathrm{d}f(\tilde{k}d)}{\mathrm{d}(\tilde{k}d)} + \frac{\mathrm{d}^2f(\tilde{k}d)}{\mathrm{d}(\tilde{k}d)^2} \right) \\
        \approx &\ \frac{6n\Omega_d}{\pi d} \int_{1/2}^{\infty} \mathrm{d}\tilde{k} \frac{\tilde{k}}{\sqrt{(2\tilde{k})^2-1}} f(\tilde{k}d) \left[\frac{1}{3\tilde{k}} \frac{\mathrm{d} f(\tilde{k}d)}{\mathrm{d}(\tilde{k}d)} + \frac{\mathrm{d}^2 f(\tilde{k}d)}{\mathrm{d}(\tilde{k}d)^2} \right] \\
        =&\ I_1(d) + I_2(d).
    \end{split}
    \end{equation}
Next,
    \begin{equation}
    \begin{split}
        I_1(d) =&\ \frac{6n\Omega_d}{\pi d} \int_{1/2}^{\infty} \mathrm{d}\tilde{k} \frac{\tilde{k}}{\sqrt{(2\tilde{k})^2-1}} f(\tilde{k}d) \frac{1}{3\tilde{k}} \frac{\mathrm{d} f(\tilde{k}d)}{\mathrm{d}(\tilde{k}d)} \\
        % =&\ \frac{2n\Omega_d}{\pi d} \int_{1/2}^{\infty} d\tilde{k} \frac{\tilde{k}}{\sqrt{(2\tilde{k})^2-1}} f(\tilde{k}d) \times \left(-\frac{1}{\tilde{k} d} \delta(x\sqrt{d} - \tilde{k})\right) \\ 
        =&\  -\frac{2n\Omega_d}{\pi d^2} \frac{x\sqrt{d}}{\sqrt{(2x\sqrt{d})^2-1}} f(x\sqrt{d})\\
        % =&\ - \frac{2 \times 2^d d d\varphi}{\pi d^2} \times \frac{x\sqrt{d}}{\sqrt{4x^2d-1}} \times \Theta(0)\\
        =&\ - \frac{2 d \tilde{\varphi}}{\pi} \times \left(1 + \frac{1}{8x^2d} + \mathcal{O}(d^{-2}) \right)
    \end{split}
    \end{equation}
and so we conclude that $I_1(d) \sim d$ at large $d$. We evaluate $I_2(d)$ next, giving 
    \begin{equation}
    \begin{split}
        I_2(d) =&\ \frac{6n\Omega_d}{\pi d} \int_{1/2}^{\infty} \mathrm{d}\tilde{k} \frac{\tilde{k}}{\sqrt{(2\tilde{k})^2-1}} f(\tilde{k}d) \times\frac{\mathrm{d}^2 f(\tilde{k}d)}{\mathrm{d}(\tilde{k}d)^2} \\
        % =&\ - \frac{6n\Omega_d}{\pi d} \int_{1/2}^{\infty} d\tilde{k} \frac{d}{d(\tilde{k}d)}\left(\frac{\tilde{k}}{\sqrt{(2\tilde{k})^2-1}} f(\tilde{k}d)\right) \times\frac{d f(\tilde{k}d)}{d(\tilde{k}d)} \\ 
        =&\ \frac{6n\Omega_d}{\pi d^3} \int_{1/2}^{\infty} \mathrm{d}\tilde{k}\left(-\frac{\tilde{k}}{\sqrt{(2\tilde{k})^2-1}} \delta(x\sqrt{d}-\tilde{k}) - \frac{\Theta(x\sqrt{d}-\tilde{k})}{((2\tilde{k})^2-1)^{3/2}}\right)\times \delta(x\sqrt{d}-\tilde{k}) \\
        =&\ -\frac{6\tilde{\varphi}}{\pi}\frac{1}{\sqrt{4x^2d-1}}\left(x\sqrt{d} \delta(0) + \frac{1}{4x^2d-1} \right),
    \end{split}
    \end{equation}
% We then evaluate the second term 
%     \begin{equation}
%     \begin{split}
%         B =&\ - \frac{6n\Omega_d}{\pi d^3} \int_{1/2}^{\infty} d\tilde{k} \frac{d}{d\tilde{k}}\left(\frac{\tilde{k}}{\sqrt{(2\tilde{k})^2-1}} f(\tilde{k}d)\right) \times\frac{d f(\tilde{k}d)}{d\tilde{k}} \\
%         =&\ \frac{6n\Omega_d}{\pi d^3} \int_{1/2}^{\infty} d\tilde{k} \frac{d}{d\tilde{k}}\left(\frac{\tilde{k}}{\sqrt{(2\tilde{k})^2-1}} f(\tilde{k}d)\right) \times \delta(x\sqrt{d}-\tilde{k}) \\ 
%         =&\ \frac{6n\Omega_d}{\pi d^3} \int_{1/2}^{\infty} d\tilde{k}\left(-\frac{\tilde{k}}{\sqrt{(2\tilde{k})^2-1}} \delta(x\sqrt{d}-\tilde{k}) - \frac{\Theta(x\sqrt{d}-\tilde{k})}{((2\tilde{k})^2-1)^{3/2}}\right)\times \delta(x\sqrt{d}-\tilde{k}) \\
%         =& - \frac{6n\Omega_d}{\pi d^3} \left( \frac{x\sqrt{d}}{\sqrt{4x^2d-1}}\delta(0) - \frac{\Theta(0)}{\left( 4x^2d-1\right)^{3/2}}\right) \\
%         =&\ -\frac{6\tilde{\varphi}}{\pi}\frac{1}{\sqrt{4x^2d-1}}\left(x\sqrt{d} \delta(0) + \frac{1}{4x^2d-1} \right) \\
%         \sim&\ d^0 + \mathcal{O}(d^{-3/2})
%     \end{split}
%     \end{equation}
where to obtain the second line we have integrated by parts. The boundary term can be shown to vanish. We then find that $I_2(d) \sim d^0 + \mathcal{O}(d^{-3/2})$ for large $d$. In turn, this allows us to conclude that $m_2 \sim d^{\lambda}$ with $\lambda=1$ (and the term is negative) for large $d$.

We can then go back to the equation for the MSD and perform a dimensional scaling analysis of the terms. We established that for the MSD, we have $\langle \Delta r^2 \rangle \sim d^{\alpha}$ with $\alpha = 1-\gamma = -5/2$. The squared cage size is given by 
    \begin{equation}
        r_s^2 = \frac{1}{2d}\langle \Delta r^2(t\rightarrow\infty)\rangle \sim d^{-1}\times d^{\alpha} = d^{-7/2}
    \end{equation}
in agreement Eq.~(17) of Jin \& Charbonneau \cite{jin2015} who performed a large $d$ analysis of the random Lorentz gas within a mode-coupling scheme.  \\

% Next we look at the scaling of the MQD $\sim d^{1+\alpha}(d^{-\gamma} + d^{\lambda})\sim d^{-5} + d^{-1/2}$. Asymptotically one would therefore keep $d^{-1/2}$ as leading behaviour although our intuition (which is true for a Gaussian process) should give $\gamma = 2\alpha$ and one would therefore keep the first term MQD$\sim d^{-5}$. 

\subsection{MCT with a Gaussian Ansatz}

Next we compute the dimensional scalings of MCT with a Gaussian ansatz for the non-ergodicity parameter (NEP). By writing $f(k) = f_s(k) = f_{\text{Gauss.}}(k) \equiv \exp[-Rk^2/2d]$, the equation of state can be written as 
    \begin{equation}
        \frac{1}{R} = \frac{n\Omega_d}{2d^2(2\pi)^d}\int_0^{\infty}\mathrm{d}k k^{d+1} c(k)^2 S(k) e^{-Rk^2/d}.
    \end{equation}
We first would like to know how $R$ scales with dimension. As before, we write $[R]\sim d^{\alpha'}$. Using same procedure as above, with the following modification: with a Gaussian ansatz, the packing fraction must be rescaled as: $\varphi = \bar{\varphi} \times 2^{-d}d.$
We write 
    \begin{equation}
    \begin{split}
        \frac{1}{R} =&\ \frac{\bar{\varphi}d}{\pi} \int_{1/2}^{\infty}\mathrm{d}\tilde{k}\frac{\tilde{k}}{\sqrt{4\tilde{k}^2-1}} e^{-Rd\tilde{k}^2}.
    \end{split}
    \end{equation}
We can define $R \equiv \bar{R}/d$ such that 
    \begin{equation}
    \begin{split}
        \frac{1}{\bar{R}} =&\ \frac{\bar{\varphi}}{\pi}\int_{1/2}^{\infty}\mathrm{d}\tilde{k}\frac{\tilde{k}}{\sqrt{4\tilde{k}^2-1}} e^{-\bar{R}\tilde{k}^2}\\
        =&\ \frac{\bar{\varphi}}{\pi} \times \frac{\sqrt{\pi}e^{-\bar{R}/4}}{4\sqrt{\bar{R}}}.
    \end{split}
    \end{equation}
This self consistent equation has finite $\bar{R}$ which is well behaved in the limit of $d\rightarrow\infty$. We therefore find that $\alpha' = -1$. From the definition of the NEP, we intuitively associate $R$ with the long-time value of the MSD and thus $\alpha'=-1$ agrees with the DMFT scaling.

\paragraph{Scaling of $m_0$}: We write in the non-ergodic phase:
    \begin{equation}
    \begin{split}
        m_0^{\text{Gauss.}} =&\ \frac{n\Omega_d}{(2\pi)^dd}\int_0^{\infty} \mathrm{d}(\tilde{k}d) (\tilde{k}d)^{d+1}c(\tilde{k}d)^2 f_{\text{Gauss.}}(\tilde{k}d)^2 \\
        =&\ \frac{2n\Omega_d}{\pi} \int_{1/2}^{\infty} d\tilde{k} \frac{\tilde{k}}{\sqrt{(2\tilde{k})^2-1}} e^{-Rd\tilde{k}^2} \\
        % =&\ \frac{2n\Omega_d}{\pi} \times \frac{1}{4}\sqrt{\frac{\pi}{\bar{R}}}e^{-\bar{R}/4} \\
        =&\ \frac{\bar{\varphi}d^2}{2\pi} \times \sqrt{\frac{\pi}{\bar{R}}}e^{-\bar{R}/4}
        % \sim&\ d^{2}.
    \end{split}
    \end{equation}
Hence, we find that $m_0^{\text{Gauss.}}\sim d^{\lambda'}$ with $\lambda' = 2$. Next, we have 
    \begin{equation}
    \begin{split}
        m_2^{\text{Gauss.}} =&\ \frac{3n \Omega_d}{(2\pi)^d d(d+2)} \int_0^{\infty}\mathrm{d}(\tilde{k}d) (\tilde{k}d)^{d+1} c(\tilde{k}d)^2f(\tilde{k}d)\left(\frac{d-1}{3(\tilde{k}d)}\frac{\mathrm{d}f(\tilde{k}d)}{\mathrm{d}(\tilde{k}d)} + \frac{\mathrm{d}^2f(\tilde{k}d)}{\mathrm{d}(\tilde{k}d)^2} \right)\\
        =&\ I_1^{\text{Gauss.}}(d) + I_2^{\text{Gauss.}}(d)
    \end{split}
    \end{equation}
We find that 
    \begin{equation}
    \begin{split}
        I_1^{\text{Gauss.}}(d) =&\ \frac{3n \Omega_d}{(2\pi)^d d(d+2)} \int_0^{\infty}\mathrm{d}(\tilde{k}d) (\tilde{k}d)^{d+1} c(\tilde{k}d)^2 e^{-Rd\tilde{k}^2} \times \frac{(d-1)R}{3d} \\
        % \approx&\ \frac{2n\Omega_d \bar{R}}{\pi d^2}\int_{1/2}^{\infty}d\tilde{k} \frac{\tilde{k}}{\sqrt{4\tilde{k}^2-1}} e^{-\bar{R}\tilde{k}^2} \\
        \approx&\ \frac{2\bar{\varphi} \bar{R}}{\pi} \times \int_{1/2}^{\infty}\mathrm{d}\tilde{k} \frac{\tilde{k}}{\sqrt{4\tilde{k}^2-1}} e^{-\bar{R}\tilde{k}^2}
    \end{split}
    \end{equation}
where in the second line we have assumed that $d\gg1$. We conclude that $I_1^{\text{Gauss.}}(d) \sim d^0$ since the integral is regular and well-behaved. An analogous calculation for $I_2^{\text{Gauss.}}(d)$ leads to the conclusion that $I_2^{\text{Gauss.}}(d) \sim d^{-1}$, and we therefore find that to leading order in $d$, $m_2 \sim I_1^{\text{Gauss.}}(d) \sim d^\eta$ with $\eta = 0$, again in agreement with the DMFT prediction discussed in the main text.

\printbibliography